\documentclass[a4paper,12pt]{article}

\usepackage{amsmath}
\usepackage{graphicx, psfrag, amsfonts, bm}
\usepackage{enumerate}
\usepackage{algorithm}
\usepackage{algpseudocode}
\usepackage{slashbox}
\usepackage{multirow}
\usepackage{mathrsfs}
\usepackage[makeroom]{cancel}
\usepackage{hhline}
\usepackage{fullpage}
\usepackage{url}
\usepackage{subcaption}
\usepackage[labelformat=simple]{subcaption}

\usepackage{natbib}
\usepackage{authblk}
\usepackage[titletoc,title]{appendix}
\RequirePackage[colorlinks,citecolor=blue,urlcolor=blue]{hyperref}
\usepackage{amssymb, pifont}
\usepackage{tabularx}

\usepackage{tikz}
\usetikzlibrary{trees}
\tikzstyle{level 1}=[level distance=0.8cm, sibling distance=0.7cm]
\tikzstyle{level 2}=[level distance=0.8cm, sibling distance=0.5cm]

\tikzstyle{bag} = [text width = 0.3em, text centered]
\tikzstyle{end} = [circle, minimum width=3pt,fill, inner sep=0pt]

\DeclareMathOperator*{\argmax}{arg\,max}

\usepackage{color}
\definecolor{brown}{rgb}{0.8, 0.33, 0.1}

\definecolor{olive}{rgb}{0.35, 0.5, 0.2}

\newcommand{\Mn}{\mbox{Mn}}
\newcommand{\Be}{\mbox{Be}}
\newcommand{\Choose}{\mbox{Choose}}
\newcommand{\Dir}{\mbox{Dir}}
\newcommand{\unif}{\mbox{Unif}}

\newcommand{\tC}{\tilde{C}}

\newcommand{\tp}{\tilde{p}}
\newcommand{\tbx}{\tilde{\bm{x}}}
\newcommand{\phat}{\hat{p}}

\newcommand{\Nbar}{\bar{N}}
\newcommand{\Tau}{\mathcal{T}}
\def\bh {\bm h}

\def\bs {\bm s}

\def\bz {\bm z}
\def\bn {\bm n}
\def\brho {\bm \rho}
\def\bx {\bm x}
\def\bZ {\bm Z}
\def\bw {\bm w}
\def\bN {\bm N}

\def\E {\text{E}}
\def\sd {\text{sd}}

\newcommand{\Chat}{\hat{C}}
\newcommand{\Tauhat}{\hat{\Tau}}
\newcommand{\Zhat}{\hat{\bZ}}
\newcommand{\what}{\hat{\bw}}

\newcommand{\MM}{\mathcal{M}}
\newcommand{\ZZ}{\mathcal{Z}}
\newcommand{\LL}{\mathcal{L}}

\newcommand{\tTau}{\tilde{\Tau}}
\newcommand{\tpois}{\text{Trunc-Pois}}
\newcommand{\err}{\text{err}}

\definecolor{grau}{rgb}{0.8,0.8,0.8}

\newcommand{\chen}[1]{\color{orange}}

\makeatletter
\renewcommand\paragraph{%
  \@startsection{paragraph}
    {4}
    {\z@}
    {3.25ex \@plus1ex \@minus.2ex}
    {-1em}
    {\normalfont\normalsize\bfseries\maybe@addperiod}%
}
\newcommand{\maybe@addperiod}[1]{%
  #1\@addpunct{.}%
}
\makeatother

\providecommand{\keywords}[1]{\textit{Keywords:  } #1}

\begin{document}

\title{TreeClone: Reconstruction of Tumor Subclone Phylogeny Based on Mutation Pairs using Next Generation Sequencing Data}
\author[1, 2]{Tianjian Zhou\thanks{Have equal contribution}}
\author[1]{Subhajit Sengupta$^*$}
\author[2, 3]{Peter M\"{u}ller\thanks{Email: pmueller@math.utexas.edu}}
\author[1, 4]{Yuan Ji\thanks{Email: yji@health.bsd.uchicago.edu }}
\affil[1]{Program for Computational Genomics and Medicine, NorthShore University HealthSystem}
\affil[2]{Department of Statistics and Data Sciences, The University of Texas at Austin}
\affil[3]{Department of Mathematics, The University of Texas at Austin}
\affil[4]{Department of Public Health Sciences, The University of Chicago}

\maketitle

\begin{abstract}
We present TreeClone, a latent feature allocation model to reconstruct tumor
subclones subject to phylogenetic evolution that mimics
tumor evolution.  Similar to most current methods, we
consider data from next-generation sequencing of tumor DNA. Unlike most methods
that use information in short reads mapped to single nucleotide
variants (SNVs), we consider subclone phylogeny reconstruction using  pairs
of  two proximal SNVs that can be mapped by the same short reads.  
As part of the Bayesian inference model, we construct a 
phylogenetic tree prior.
The use of the tree structure in the prior greatly strengthens
inference. Only subclones that can be explained by a
phylogenetic tree are assigned non-negligible probabilities. 
The proposed Bayesian framework  implies  posterior
distributions on the number of subclones, their genotypes, 
cellular proportions, and the phylogenetic tree spanned by the
inferred subclones. The proposed method is validated against
different sets of simulated and real-world data using single and
multiple tumor samples.  An open source software package is 
available at \url{http://www.compgenome.org/treeclone}.
\end{abstract}

\noindent\keywords{
Latent feature model; Mutation pair; NGS data; Phylogenetic tree; Subclone; Tumor heterogeneity}

\section{Introduction}
\label{sec:intro}

Initiated by somatic mutations in a single cell, cancer arises through Darwinian-like natural selection. The accumulation of subsequent genetic aberrations and the effects of selection over time result in the sequential clonal expansions of cells, finally leading to the development of a genetically aberrant
tumor~[\cite{nowell1976clonal}]. This process, known as tumorigenesis,
leads to genetically divergent subpopulations of tumor cells, also known as subclones~[\cite{bonavia2011heterogeneity,marusyk2012intra}]. 

Deep next generation sequencing (NGS) of bulk tumor DNA samples makes it
possible to examine the evolutionary history of individual tumors,
based on the set of somatic mutations they have
accumulated~[\cite{nik2012life}].
This is implemented by deconvoluting
observed genomic data from a tumor into constituent signals
corresponding to various subclones and  by then reconstructing 
the relationship of these subclones 
in a phylogeny~[\cite{deshwar2015phylowgs,
    marass2017phylogenetic}]. Apart from phylogenetic relationship,
tumor purity, subclones' genotypes and cellular proportions are also 
coupled quantities to infer. Uncovering subclonal heterogeneity
and their relationship is clinically important for better
prognosis~[\cite{aparicio2013implications,
    schwarz2015spatial}]. Therefore there is a pressing need to
develop robust methods for a reliable interpretation. 

Numerous methods have been proposed for the subclonal reconstruction
problem,
including
SciClone [\cite{miller2014sciclone}],
CloneHD [\cite{Fischer:2014}],
PyClone [\cite{roth2014pyclone}],
PyloWGS [\cite{jiao2014inferring,deshwar2015phylowgs}],
Clomial [\cite{zare2014inferring}],
BayClone [\cite{sengupta2015bayclone}],
Cloe [\cite{marass2017phylogenetic}],
and PairClone [\cite{zhou2017pairclone}].
The reconstruction is typically based on
short reads that are mapped to single nucleotide variants
(SNVs) (few methods, for example, CloneHD, also consider somatic copy number aberrations, SCNA).  
 Many methods based on SNV data  utilize variant allele fractions
(VAFs), that is, the fractions of alleles (or short reads) at each locus that carry mutations. 
Since humans are diploid, the expected VAFs of short reads for a
homogeneous cell population should be 0, 0.5 or 1.0 for any locus in
copy number neutral (copy number = 2) regions and after adjusting for
tumor purity. Observing VAFs very different from 0, 0.5 or 1.0 is
therefore evidence for heterogeneity.  Most methods use only marginal
SNVs.
Recently, \cite{zhou2017pairclone} have proposed to use pairs of
proximal SNVs mapped by the same short reads, which carry more
information than marginal SNVs, for more accurate subclone
reconstruction. 
In terms of methodology, existing subclone reconstruction methods can
be mainly divided into two categories: clustering-based and
feature-allocation-based. The two categories are also referred to as
indirect and direct reconstructions in \cite{marass2017phylogenetic},
depending on whether the subclonal genotypes are indirectly or
directly inferred.  Clustering-based methods,
 including SciClone, PyClone and PhyloWGS, 
first infer SNV clusters
based on observed VAFs and then reconstruct subclonal genotypes based
on the clusters. The phylogenetic relationship among the subclones can
be inferred by imposing hierarchy among the SNV clusters.
On the other hand, feature-allocation-based methods
(e.g., Clomial, BayClone, Cloe or PairClone)
treat subclones as latent features and directly infer subclonal
genotypes. Most of the feature-allocation-based methods
assume that the features (subclones) are conditionally independent and thus
are not able to infer the phylogenetic relationship among the
subclones. Recently, \cite{marass2017phylogenetic} have developed a
model allowing for dependency among the features to infer the tumor
phylogenetic tree. 

\begin{figure}[h!]
\centering
\includegraphics[width=\textwidth]{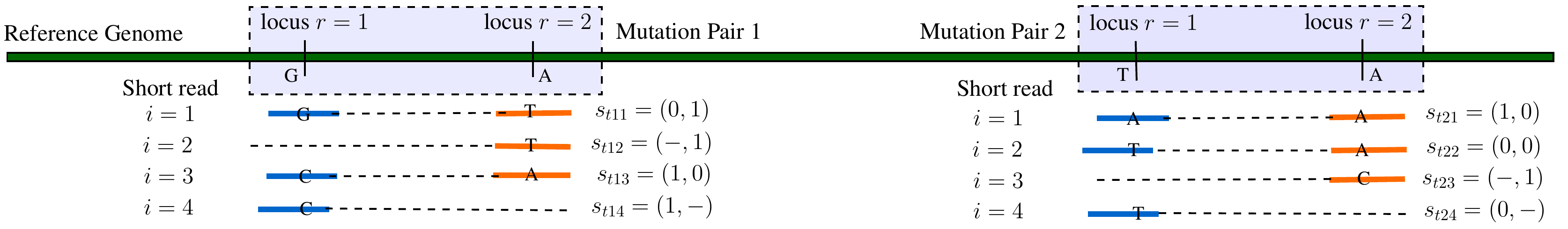}
\caption{Short reads data from mutation pairs
  using NGS. Here $s_{tki}$ denotes the $i$-th read
  for  the $k$-th mutation pair in sample $t$. 
  Each $s_{tki}$ is a 2-dimensional vector  which corresponds to the
  two proximal SNVs in the mutation pair, and each component of the
  vector takes values $0$, $1$ or --, representing wild type,
  variant or missing genotype, respectively.} 
\label{fig:lochap_data}
\end{figure}

 In the upcoming discussion we 
assume that the available data are from $T$ ($T\geq1$) samples 
from a single patient and the main inference goal is
intra-tumor heterogeneity.
We present a novel approach to reconstruct tumor
subclones and their corresponding phylogenetic tree based on
mutation pairs. Here a mutation pair refers to a pair of proximal
SNVs on the genomes that can be simultaneously mapped by the same
paired-end short reads, with one SNV on each end.
In other words, mutation pairs can be phased by short reads. See
Fig. \ref{fig:lochap_data} for an illustration. 
Short reads mapped to only one of the SNV loci are treated as
partially missing paired-end reads and are not excluded from our
approach.
This idea of working with phased mutation pairs was introduced  in \cite{zhou2017pairclone}.
We build on this work and develop a novel and entirely
different inference approach by explicitly modeling the
underlying phylogenetic relationship.
That is, we model tumor heterogeneity based on a representation of
a phylogenetic tree of tumor cell subpopulations. A prior probability
model on such phylogenetic trees induces a dependent prior on
the mutation profiles of latent tumor cell subpopulations.
 Part of this construction is that  the phylogenetic tree of
tumor cell subpopulations is included as a 
random quantity in the Bayesian model.  Like most existing methods, we
only consider 
mutation pairs in copy number neutral region i.e. copy number two.
The proposed inference aims
to reconstruct (i) subclones defined by the haplotypes across all the
mutation pairs, (ii) cellular proportion of each subclone, and (iii) a
phylogenetic tree spanned by the subclones.

 Next we introduce some notation to formally represent the
described data and model structure. 
Consider an NGS data set with $K$ mutation pairs shared
across all $T$ ($T \ge 1$) samples.
We assume that the samples are composed of $C$
homogeneous subclones. The number of subclones $C$ is unknown and
becomes one of the model parameters.
We use a $K \times C$ matrix $\bZ$ to represent
the subclones, in which each column of $\bZ$ represents a subclone and
each row represents a mutation pair. 
That is, the $(kc)$ element  $\bz_{kc}$ of the matrix corresponds to
subclone $c$ and mutation pair $k$.
Each $\bz_{kc}$ is itself again a matrix. It is a $2 \times 
2$ matrix that represents the genotypes of the two alleles of the
mutation pair. See Fig. \ref{fig:sc_evo}(b).
An important step in the model construction is that 
the columns (subclones) of $\bZ$ form a phylogenetic tree $\Tau$.
The tree encodes the parent-child relationship across the subclones. 
A detailed construction of the tree and a prior probability model of
$\Tau$ and $\bZ$ are introduced later.
Lastly, we denote with $\bw_t = (w_{t1}, \ldots, w_{tC})$ the cellular
proportions of the subclones in sample $t$ where $0 < w_{tc} < 1$ for
all $c$ and $\sum_{c=0}^C w_{tc} = 1$.  
Using NGS data we infer $\Tau$, $C$, $\bZ$ and $\bw$ based on a simple
idea that variant reads can only arise from subclones with variant
alleles consistent with an underlying phylogeny. We develop a tree-based
{\em latent feature allocation model} (LFAM) to implement this
reconstruction.  Mutation pairs are the objects of the LFAM, and
subclones are the latent features chosen by the mutation pairs.

 The previous brief outline of data and model structure already
highlights two key features of the proposed approach: the use of
phylogenetic tree structure and data on mutation pairs.
The latter has important advantages. 
Mutation pairs contain phasing information that improves the accuracy
of subclone reconstruction.  If two nucleotides reside on the same
short read, we know that they must appear in the same DNA strand in a
subclone. For example, consider a scenario with one mutation pair and
two subclones. Suppose the reference genome allele is (A, G) for that
mutation pair, with the notion that A and G are phased by the same DNA
strand.  Suppose the two subclones have diploid genomes at the two
loci and the genotypes for both DNA strands are ((C, G), (A, T)) for
subclone $c=1$, and ((C, T), (A, G)) for $c=2$.  Since in NGS short
reads are generated from a single DNA strand, short reads could be any
of the four haplotypes (C, G), (A, T), (C, T) or (A, G) for this
mutation pair.  If indeed relative large counts of short reads with
each haplotype are observed, one can reliably infer that there are
heterogeneous cell subpopulations in the tumor sample and the mutation pairs are subclonal.
In contrast, if we ignore the phasing information and only consider
the (marginal) VAFs for each SNV, then the observed VAFs for both SNVs
are 0.5, which could be explained by heterogeneous clonal mutations, i.e., 
the SNVs are present in all tumor cells.
In this paper, we leverage the power of using mutation pairs over single
SNVs to incorporate partial phasing information in our model. 
We assume that mutation pairs and their mapped short reads counts have
been obtained using a bioinformatics pipeline, such as 
\texttt{LocHap}
[\cite{sengupta2016ultra}]. Our aim is to use short reads mapping data
on mutation pairs to reconstruct tumor subclones and their
phylogeny. 

 Besides the use of mutation pairs, the other key feature of the
proposed approach is that the model is built around
phylogenetic tree structure. 
Imposing the phylogenetic tree structure in the prior of $\bZ$ greatly
strengthens inference. First, the tree structure improves biological
interpretability of the inferred subclones as the evolutionary
relationship among the subclones is explicitly modeled. Second, the
tree structure improves the identifiability of the problem. In a
subclone reconstruction problem, the input signals (observed VAFs) are
usually relatively weak, especially when only $T=1$ sample is
available. Different subclone architectures can yield very similar
observed data.  
By explicitly modeling the tree we can put higher prior probability on
a subclone structure that follows a more likely phylogenetic tree.
Third, the tree structure improves the mixing performance of the
Markov chain Monte Carlo (MCMC) simulation used to infer the unknown
quantities. As noted in \cite{marass2017phylogenetic}, the
 likelihood 
surface of the genotype matrix $\bZ$ is highly multi-modal with sharp
peaks. Imposing the tree structure, in the MCMC simulation we only
need to search from the space of $\bZ$ where the tree structure is
satisfied, which greatly reduces the dimension of the parameter space
of $\bZ$ thus improves mixing of the Markov chain.

Finally, for clarification we briefly comment on the proposed model
structure versus a traditional use of phylogenetic trees.
Phylogenetic trees are usually used to approximate perfect phylogeny
for a fixed number of
haplotypes
[\cite{bafna2003haplotyping}].
Most methods lack assessment of tree uncertainties and report a
single tree estimate. Also,  methods based on SNVs 
put the observed mutation profile of SNV at the leaf nodes.
This is natural if the splits in the tree create subpopulations
that acquire or do not acquire a new mutation (or set of mutations).
In contrast, we define a tree with all descendant nodes differing from
the parent node by some mutations. 
That is, all nodes, including interior nodes, correspond to
tumor cell subpopulations.
Finally, we note that 
the prior structure in our model is different from the phylogenetic
Indian Buffet Process (pIBP) [\cite{miller2012phylogenetic}], which
models phylogeny of the objects rather than the features.\\

The rest of the paper is organized as follows: Section~\ref{sec:model}
and Section~\ref{sec:post} describe the latent feature allocation
model and posterior inference, respectively. Section~\ref{sec:sim}
presents two simulation studies. Section~\ref{sec:real} reports
analysis results for an actual experiment.  We conclude with a
discussion in Section~\ref{sec:disc}.

\section{Statistical Model}
\label{sec:model}
\subsection{Representation of Subclones}
\label{sec:rep_subclone}
\begin{figure}[h!]
\centering
\begin{subfigure}[t]{.53\textwidth}
\centering
\includegraphics[width=\textwidth]{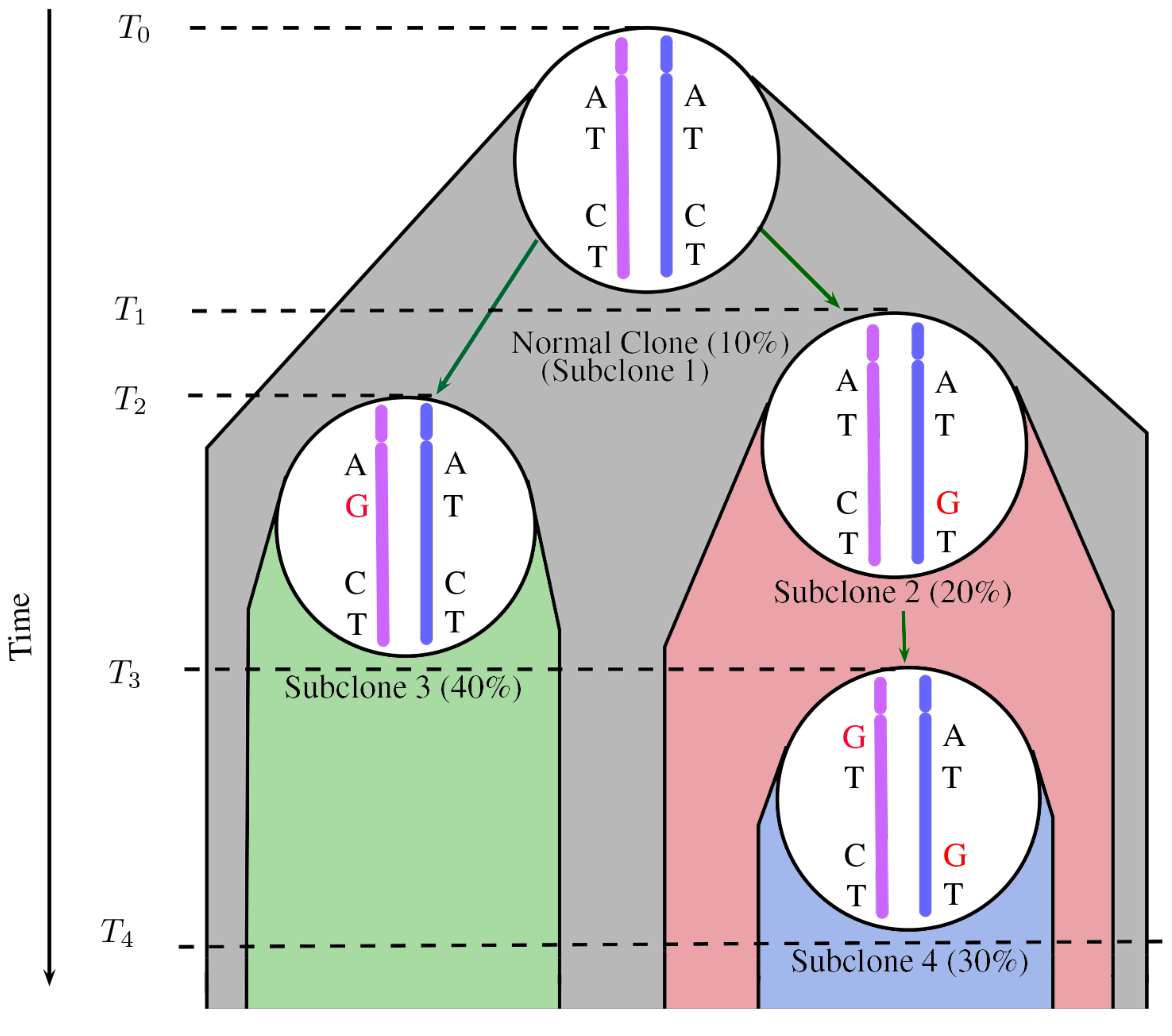}
\caption{}
\end{subfigure}
\begin{subfigure}[t]{.46\textwidth}
\centering
\includegraphics[width=\textwidth]{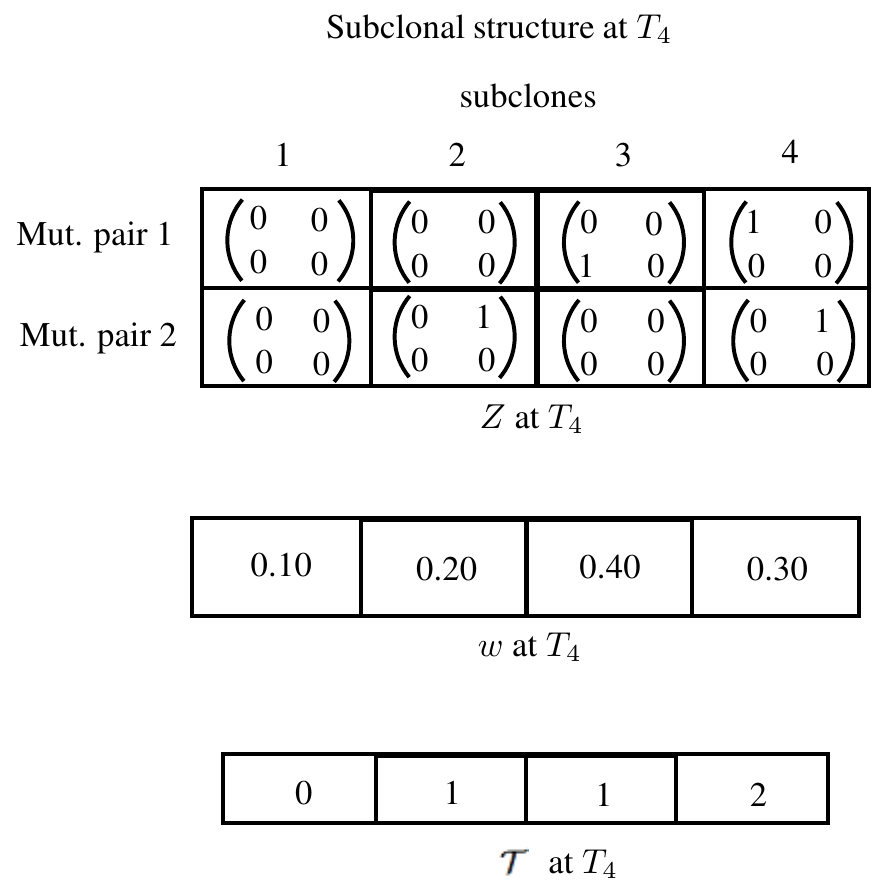}
\caption{}
\end{subfigure}
\caption{Schematic of subclonal evolution and subclone
  structure. Panel (a) shows the evolution of subclones over time.
  Panel (b) shows the subclonal structure at $T_4$ with 
  genotypes $\bZ$, cellular proportions $\bw$ and parent vector $\Tau$. 
  For each mutation pair $k$ and subclone $c$, 
  the entry $\bz_{kc}$ of $\bZ$ is a $2\times2$ matrix corresponding to
  the arrangement in the figure in panel (a), that is, 
  with alleles in the two columns, and SNVs in the rows.}
\label{fig:sc_evo}
\end{figure}

Fig.~\ref{fig:sc_evo} presents a stylized example of temporal
evolution of a tumor, starting from time $T_0$ and evolving until time
$T_4$ with the normal clone (subclone $c=1$) and three tumor subclones
($c=2,3,4$).
Each tumor subclone is marked by two mutation pairs with distinct
somatic mutation profiles. In Fig.~\ref{fig:sc_evo} the true
phylogenetic tree is plotted connecting the stylized subclones.  The
true population frequencies of the subclones are marked in
parentheses. In panel (b) subclone genomes, their
population frequencies and the phylogenetic relationship are
represented by $\bZ$, $\bw$, and $\Tau$.
 Next we explain in detail the definition of these parameters. 

The entries of $\Tau$ 
report for each subclone the index of the parent subclone
(with $\Tau_1=0$ for the root clone $c=1$).
Suppose $K$ mutation pairs with $C$ subclones are
present. The subclone phylogeny can be visualized with a rooted tree
with $C$ nodes.  We use a $C$-dimensional {\it parent} vector $\Tau$ 
to encode the parent-child relationship of a
tree, where $\Tau_c = \Tau[c] = j$ means that subclone $j$ is
the parent of subclone $c$. The parent vector uniquely
defines the topology of a rooted tree.  We assume that the tumor
evolutionary process always starts from the normal clone, indexed
by $c = 1$. The normal clone does not have a parent, and we denote
it by $\Tau_1 = 0$. For example, the parent vector representation of
the subclone phylogeny in Fig. \ref{fig:sc_evo} is $\Tau = (0, 1, 1,
2)$.

We use the $K \times C$ matrix $\bZ$ to represent the subclone
genotypes. Each column of $\bZ$ defines a subclone, and each row of
$\bZ$ corresponds to a mutation pair.  The entry $\bz_{kc}$ records
the genotypes for mutation pair $k$ in subclone $c$.  Since each
subclone has two alleles $j = 1, 2$, and each mutation pair has two
loci $r = 1, 2$, the entry $\bz_{kc}$ is itself a $2 \times 2$ matrix,
\begin{equation*}
\bz_{kc} = (\bz_{kc1}, \bz_{kc2}) =  
\begin{bmatrix}
   \left(\begin{array}{c}
        z_{kc11}  \\
        z_{kc12}  \\
  \end{array}\right) 
   \left(\begin{array}{c}
        z_{kc21}  \\
        z_{kc22}  \\
  \end{array}\right) 
\end{bmatrix}
\end{equation*}
where 
$\left(\begin{array}{c} 
         z_{kc11}  \\ z_{kc12}  \\ 
 \end{array}\right)$ 
and 
$\left(\begin{array}{c} 
         z_{kc21}  \\ z_{kc22}  \\ 
\end{array}\right)$ 
represent mutation pairs of allele $1$ and allele $2$, respectively.
 That is, in $z_{kcjr}$, $j$ and $r$ index the two alleles and the
two loci, respectively. 
Theoretically, each $z_{kcjr}$ can be any one of the four nucleotide
sequences, A, C, G, T. However, at a single locus, the probability of
having more than two sequences is negligible since it would require
the same locus to be mutated twice throughout the life span of the
tumor, which is extremely unlikely.
Therefore, we assume $z_{kcjr}$ can only take two possible values,
with $z_{kcjr} = 1$ (or $0$) indicating that the corresponding locus has a
mutation (or does not have a mutation) compared to the reference
genome, respectively.  
For example, in Fig. \ref{fig:sc_evo}, we have $K = 2$ mutation pairs
and $C = 4$ subclones.
For mutation pair $k = 2$ in subclone $c = 4$, the allele $j=1$
harbors no mutation, while the allele $j=2$ has a mutation at the
first locus $r=1$, which translates to $\bz_{24} = (00, 10)$ (writing
00 as a shorthand for $(0,0)^T$, etc.). 
Altogether, $\bz_{kc}$ can take $2^4 = 16$ possible
values $\bz_{kc} \in \{(00, 00), (00, 01), \ldots, (11,11) \}$. Since
we do not have phasing information across mutation pairs, the
$\bz_{kc}$ values having mirrored columns  lead to exactly the same data
likelihood and thus are indistinguishable. 
 This reduces the list of possible values of $\bz_{kc}$ to the
$Q=10$ values, 
$\bz^{(1)} = (00, 00)$, $\bz^{(2)} = (00, 01)$,
$\bz^{(3)} = (00, 10)$, $\bz^{(4)} = (00, 11)$, $\bz^{(5)} = (01,
01)$, $\bz^{(6)} = (01, 10)$, $\bz^{(7)} = (10, 10)$, $\bz^{(8)} =
(01, 11)$, $\bz^{(9)} = (10, 11)$ and $\bz^{(10)} = (11, 11)$. \\
We assume that the normal subclone has no mutation, $\bz_{k1} = \bz^{(1)}$ for
all $k$, indicating all mutations are somatic. 

Finally, we introduce notation for mixing proportions $\bw$.
Suppose $T$ tissue samples are dissected from the same patient. We
assume that the samples are admixtures of $C$ subclones, each sample
with a different set of mixing proportions (population
frequencies). We use a  $T \times C$  matrix $\bw$ to record the
proportions, where $w_{tc}$ represents the population frequencies
of subclone $c$ in sample $t$, $0 < w_{tc} < 1$ and $\sum_{c=1}^C
w_{tc} = 1$. The proportions $w_{t1}$ denotes the proportion of
normal cells contamination in sample $t$
 (and later we will still add a weight $w_{t0}$ for a background
clone $c=0$). 

\subsection{Sampling Model}
\label{sc:sampling_model}

Let $\bN$ be a $T \times K$ matrix with $N_{tk}$ representing read
depth for mutation pair $k$ in sample $t$. It records the number of
times any locus of the mutation pair is covered by sequencing reads
(see Fig.~\ref{fig:lochap_data}). 
Let $\bs_{tki} = \left(s_{tkir}, r = 1, 2 \right)$ be a specific short
read where $r = 1, 2$ index the two loci in a mutation pair, $i = 1,
2, \ldots, N_{tk}$. We use $s_{tkir} = 1$ (or $0$) to denote a
variant (reference) sequence at the read, 
compared to the reference genome. An important feature of the data is
that read $i$ may   overlap in only one locus. 
We use $s_{tkir} = -$ to represent the missing  other  sequence on the
read. Reads that do not overlap with either of the two loci are not
included in the model as they do not contribute any information about
the mutation pair. In summary, $\bs_{tki}$ can take $G = 8$
possible values, 
$$
   \bs_{tki} \in \{\bs^{(1)}, \ldots, \bs^{(8)} \} =
       \{00, 01, 10, 11, -0, -1, 0-, 1- \}.
$$
Among all $N_{tk}$ reads, let
$n_{tkg} = \sum_i I \left(\bs_{tki} = \bs^{(g)} \right)$ be the number
of short reads having genotype $\bs^{(g)}$. 
 E.g., in  Fig.~\ref{fig:lochap_data} out of a total of
$N_{t1}=4$ reads, we 
have $n_{t12} = 1, n_{t13} = 1,n_{t16} = 1$ and $n_{t18}=1$.

We assume a multinomial sampling model for the observed read counts
\begin{align}
   (n_{tk1}, \ldots, n_{tk8}) \mid N_{tk} \sim \Mn(N_{tk}; p_{tk1}, \ldots, p_{tk8}),
\label{eq:multi-sampling}
\end{align}
where $p_{tkg}$ is the probability of observing a short read $\bs_{tki}$
with genotype $\bs^{(g)}$.
 Next  we link $p_{tkg}$ with the
underlying subclone structures. 

For a short read $\bs_{tki}$, depending on whether it covers both loci
or only one locus, we consider three cases: (i) a read covers both
loci, taking values $\bs_{tki} \in \{\bs^{(1)}, \ldots, \bs^{(4)} \}$
(complete read); (ii) a read covers the second locus, taking values
$\bs_{tki} \in \{\bs^{(5)}, \bs^{(6)} \}$ (left missing read); and
(iii) a read covers the first locus, taking values $\bs_{tki} \in
\{\bs^{(7)}, \bs^{(8)} \}$ (right missing read).  Let $v_{tk1},
v_{tk2}, v_{tk3}$ denote the probabilities of observing a short read
 of type  (i), (ii) and (iii), respectively. Conditional on
cases (i), (ii) or (iii), let $\tp_{tkg}$ be the conditional
probability of observing $\bs_{tki} = \bs^{(g)}$. We have $p_{tkg} =
v_{tk1} \, \tp_{tkg}, g = 1, \ldots, 4$, $p_{tkg} = v_{tk2} \,
\tp_{tkg}, g = 5, 6$, and $p_{tkg} = v_{tk3} \, \tp_{tkg}, g = 7,
8$. We assume non-informative missingness and do not make inference on
$v$'s. So they remain constants in the likelihood.

We express $\tp_{tkg}$ in terms of $\bZ$ and $\bw$ based on the
following generative model in three steps.
Consider a sample $t$. To generate a
short read, we first select a subclone $c$ with probability $w_{tc}$
 (step 1). 
Next we select with probability $0.5$ one of the two alleles $j= 1 ,
2$ (step 2). Finally, we record the read $\bs^{(g)}$, $g=1,2,3$ or $4$,
corresponding to the chosen allele $\bz_{kcj}=(z_{kcj1},z_{kcj2})$
(step 3).
In the case of left (or right) missing locus we observe $\bs^{(g)}$,
$g=5$ or $6$ (or $g=7$ or $8$), corresponding to the observed locus of
the chosen allele.
Reflecting  steps 2 and 3,  we denote the probability of
observing a short read $\bs^{(g)}$ from subclone
 characterized by $\bz_{kc}$ by 
\begin{equation}
 A(\bs^{(g)}, \bz_{kc}) =
   \sum_{j = 1}^2 0.5\,\times\,I(s^{(g)}_1 = z_{kcj1})\, I(s^{(g)}_2=z_{kcj2}),
 \label{eq:A}
\end{equation}
with the understanding that $I(- = z_{kcjr}) \equiv 1$ for missing reads.
Implicit in \eqref{eq:A} is the restriction
$A(\bs^{(g)}, \bz_{kc}) \in \{0, 0.5, 1\}$, depending on the arguments.
Finally,  proceeding as in step 1 we use 
the conditional probabilities $A(\cdot)$ to obtain the marginal probability
of observing a short read $\bs^{(g)}$ from the tumor sample $t$ with
$C$ subclones with cellular proportions $\{ w_{tc}\}$ as 
\begin{equation}
   \tp_{tkg} = \sum_{c = 1}^C w_{tc}\,A(\bs^{(g)}, \bz_{kc}) + w_{t0} \, \rho_g.
   \label{eq:ptkg}
\end{equation}
The first term in Eq.~\ref{eq:ptkg} states that the probability of
observing a short read with genotype $\bs^{(g)}$ is a weighted sum of
the $A$'s across all the subclones.

The last term introduces the notion of a background subclone,
indexed as $c=0$ and without biological meaning, to account
for noise and artifacts in the NGS data
(sequencing errors, mapping errors, etc.), and also for tiny subclones
that are not detectable given the sequencing depth.
In \eqref{eq:ptkg}, $w_{t0} \rho_g$ stands for the probability of
observing $\bs^{(g)}$ due to this random noise.
We assume the random noise does not differ across different mutation
pairs, thus $\rho_g$ does not have an index $k$. Note that
$\rho_1+\ldots + \rho_4= \rho_5+\rho_6 = \rho_7+\rho_8 = 1$. 

 Finally, we note that 
if desired, it is straightforward to incorporate data for marginal SNV
reads in the sampling model by treating such reads as, for example, 
right missing reads, i.e. $s_{tki2} = -$. In this case,
$n_{tk1} = \ldots = n_{tk6} = 0$, and the multinomial sampling model
reduces to a binomial model. The addition of marginal SNV counts does
not typically improve inference. See more details in
\cite{zhou2017pairclone}.

\subsection{Prior Model}
We construct a hierarchical prior model, starting
with $p(C)$, then a prior on the tree for a given number of nodes,
$p(\Tau \mid C)$, and  finally a prior on the subclonal genotypes
given the phylogenetic tree $\Tau$.
We assume a geometric prior for the number of subclones,
$p(C) = (1 - \alpha)^{C - 1} \alpha$, $C \in \{1, 2, 3,
\ldots\}$. Conditional on $C$, the prior on the tree, 
$p(\Tau \mid C)$ is as in~\cite{chipman1998bayesian}. For a tree with $C$ nodes, we let 
\begin{align*}
p(\Tau \mid C) \propto \prod_{c = 1}^C (1 + \eta_c)^{-\beta}, 
\end{align*}
where $\eta_c$
is the depth of node $c$, or the number of generations between node
$c$ and the normal subclone $1$. The prior penalizes deeper trees and
thus favors parsimonious representation of subclonal structure. 


 Conditional on $\Tau$ we define a prior for $\bZ$. 
The subclone genotype matrix $\bZ$ can be thought of as a feature
allocation for categorical matrices. The mutation pairs are the
objects, and the subclones are the latent features chosen by the
objects. Each feature has 10 categories corresponding to the $Q=10$ different
genotypes. Given $\Tau$ the 
construction of the subclone genotype matrix needs to introduce
dependence across features to respect the assumed phylogeny.
We construct a prior for $\bZ$ based on the following
generative model.  
We start from a normal subclone denoted by $\bz_{\cdot 1} = \bm 0$. 
Now consider a subclone $c>1$ and defined by $\bz_{\cdot c}$. 
The subclone preserves all mutations from its parent $\bz_{\cdot \Tau_c}$,
but also gains a Poisson number of new mutations. We assume the new
mutations happen randomly at the unmutated loci of the parent
subclone. A formal description of prior of $\bZ$ follows.

For a subclone $c$, let $\ell_{kc} = \sum_{j, r} z_{kcjr}$ denote the
number of mutations in mutation pair $k$, and let
$\mathcal{L}_c = \{k: \ell_{kc} < 4 \}$ denote the mutation pairs
in subclone $c$ that have less than four mutations. 
Let $m_{kc} = \ell_{kc} - \ell_{k \Tau_c}$ denote the number of new mutations that
mutation pair $k$ gains compared to its parent, and let $m_{\cdot c} =
\sum_k m_{kc}$. 
We assume
(i) The child subclone should acquire at least one additional mutation
compared with its parent (otherwise subclone $c$ would be identical
to its parent $\Tau_c$).
(ii) If the parent has already acquired all four mutations for a
given $k$, then the child can not gain any more new mutation.   
That is, if $\ell_{k \Tau_c} = 4$, then $m_{kc} = 0$.  
(iii) Each mutation pair can gain at most one additional mutation in
each generation, $m_{kc} \in \{0, 1\}$.  
Based on these assumptions,  given a parent subclone
$\bz_{\cdot \Tau_c}$, we construct a child subclone $\bz_{\cdot c}$ as follows. 
Let $\MM_c = \{k: m_{kc} = 1 \}$ be the set of   mutation   pairs in
subclone $c$  where new mutations are gained.
Let $\Choose(\LL, m)$ denote
a uniformly chosen subset of $\LL$ of size $m$, and let
$X \sim \tpois(\lambda; [a, b])$ represent a Poisson distribution with mean
$\lambda$, truncated to $a \leq X \leq b$. We assume
\begin{align}
 m_{\cdot c} \mid \bz_{.\Tau_c},\Tau,C &\sim \tpois(\lambda; [1, |\mathcal{L}_{\Tau_c}| ]), \nonumber\\
  \MM_c \mid m_{.c},\bz_{.\Tau_c},\Tau,C &\sim \Choose(\LL_{\Tau_c}, m_{\cdot c}).
\label{eq:prior_z}
\end{align}
The lower bound and upper bound of the truncated Poisson reflect
assumptions (i) and (ii) respectively. 
Also, Eq.~\ref{eq:prior_z} implicitly captures assumption (iii).    

Next, for a mutation pair that gains one new mutation, we assume the
new mutation randomly arises in any of the unmutated loci in the
parent subclone.
Let $\ZZ_{kc} = \{(j, r): z_{kcjr} = 0 \}$,  and let 
$\unif(A)$ denote a uniform distribution over the set $A$.
We first choose
\begin{align*}
(j^*, r^*) \mid \bz_{.\Tau_c},\Tau,C \sim \unif(\ZZ_{k \Tau_c}),
\end{align*}
and then set $z_{kcj^* r^*} = 1$. 
So we have 
\begin{align*}
p(\bZ \mid \Tau, C) \propto \prod_{c=2}^C \tpois(m_{\cdot c}; [1, |\mathcal{L}_{\Tau_c}| ]) . \frac{1}{\left(\begin{array}{c} |\mathcal{L}_{\Tau_c}| \\ m_{\cdot c} \\ 
 \end{array}\right)} . \prod_{k\in \mathcal{M}_c} \frac{1}{|\ZZ_{k \Tau_c}|}.
\end{align*}

 The prior construction is completed with a
prior for $\bw$ and $\brho$. 
We design the prior $p(\bw)$ of $\bw$ in such a manner that we could put an informative prior for
$w_{t1}$ if a reliable estimate for tumor purity is available based
on some prior bioinformatics pipeline 
(e.g.~\cite{van2010allele,carter2012absolute}). 
Recall that $c=1$ is the normal subclone, i.e.,
$w_{t1}$ is the normal subclone proportion (or 1 minus tumor purity), and
that $\sum_{{c=0},{c\neq 1}}^C w_{tc} + w_{t1} = 1$. We assume a
Beta-Dirichlet prior on $\bw$ such that,
\begin{equation*}
  w_{t1} \sim \Be(a_p,b_p);\quad \mbox{and} \;\;
  \frac{w_{tc}}{(1-w_{t1})} \sim \Dir (d_0,d,\cdots,d),
\end{equation*}
where $c = 0, 2, 3,\cdots,C$. 
We set $d_0 << d$ as $w_{t0}$ is only a correction term to account for background noise and model mis-specification term.

 Finally, to specify a prior  for $\brho = \{\rho_g\}$ we
consider complete reads, left missing reads and right missing reads
separately and assume
$(\rho_1,\ldots,\rho_4) \sim \Dir(d_1, \ldots, d_1)$,
$(\rho_5,\rho_6) \sim  \Dir(2d_1, 2d_1)$, and
$(\rho_7,\rho_8) \sim  \Dir(2d_1, 2d_1)$.

\section{Posterior Inference}
\label{sec:post}
Let $\bx = (\bZ, \bw, \brho)$ denote the unknown parameters except for
the number of subclones $C$ and the tree $\Tau$. Markov chain Monte
Carlo (MCMC) simulation from the posterior $p(\bx
\mid \bn, \Tau, C)$ is used to implement posterior inference.
Gibbs sampling transition probabilities are used
to update $\bZ$, and Metropolis-Hastings transition probabilities are
used to update $\bw$ and $\brho$. For example, we update $\bZ$ by row
with 
\begin{align*}
p(\bz_{k \cdot} \mid \bz_{-k \cdot}, \ldots) \propto \prod_{t = 1}^T \prod_{g = 1}^G \left[ \sum_{c = 1}^C w_{tc} \, A(\bh_g, \bz_{kc}) + w_{t0} \, \rho_g \right]^{n_{tkg}} \cdot 
p(\bz_{k \cdot} \mid  \bz_{-k \cdot}, \Tau, C),
\end{align*}
where $\bz_{k \cdot}$ is a row of $\bZ$ satisfying the phylogeny
$\Tau$.  

Since the posterior distribution $p(\bx \mid \bn, \Tau, C)$
is expected to be highly multi-modal, we utilize parallel
tempering [\cite{geyer1991markov}] to improve the mixing of the
chain. Specifically, we use OpenMP parallel computing
API [\cite{dagum1998openmp}] in C++, to implement a scalable parallel
tempering algorithm.

Updating $C$ and $\Tau$  is more difficult. 
In general, posterior MCMC on tree structures can be very
challenging to implement [\cite{chipman1998bayesian,Deni:Mall:Smit:baye:1998}].
However, the problem here is manageable since plausible numbers for
$C$ constrain $\Tau$ to moderately small trees.
We assume that the number of nodes is {\it a priori} restricted to
$C_{\min} \leq C \leq C_{\max}$. Conditional on the number of
subclones $C$, the number of possible tree topologies is
finite.
Let $\mathscr{T}$ denote the (discrete) sample space of $(\Tau, C)$. 
Updating the values of $(\Tau, C)$ involves trans-dimensional MCMC.
At each iteration, we propose new values for $(\Tau, C)$ from a
uniform proposal, $q(\tTau, \tC \mid \Tau, C) \sim
\unif(\mathscr{T})$.

In order to search the space $\mathscr{T}$ for the number of subclones
and trees that best explain the observed data, we follow a
similar approach as in \cite{lee2015bayesian, zhou2017pairclone} (motivated by
fractional Bayes' factor in \cite{OHagan:95}) that splits the data
into a training set and a test set. Recall that $\bn$ represents the
read counts data. We split $\bn$ into a training set
$\bn'$ with $n_{tkg}' = b n_{tkg}$, and a test set $\bn''$ with
$n_{tkg}'' = (1-b) n_{tkg}$.  
Let $p_b(\bx \mid \Tau, C) = p(\bx \mid
\bn', \Tau, C)$ be the posterior evaluated on the
training set only. We use $p_b$ in two instances. First, $p_b$ is used
as an informative prior instead of the original prior $p((\bx \mid
\Tau, C)$, and second, $p_b$ is used as a proposal distribution for
$\tbx$, $q(\tbx \mid \tilde{\Tau}, \tC) = p_b(\tbx \mid \tilde{\Tau},
\tC)$. Finally, the acceptance probability of proposal $(\tilde{\Tau},
\tC, \tbx)$ is evaluated on the test set.
Importantly, in the acceptance probability the (intractable) normalization
constant of $p_b$ cancels out, making this approach computationally
feasible.
\begin{multline*}
  p_{\text{acc}}(\Tau, C, \bx, \tTau, \tC,\tbx) = 1 \wedge
  \frac{p(\bn'' \mid \tbx, \tTau, \tC)}
       {p(\bn'' \mid \bx, \Tau, C)} \cdot 
  \frac{p(\tTau, \tC) \cancel{p_b(\tbx \mid \tTau, \tC)}}
       {p(\Tau, C)   \cancel{p_b(\bx  \mid \Tau, C  )}} \cdot \\
  \frac{q(\Tau, C \mid \tTau, \tC) \cancel{q(\bx \mid \Tau, C)}}
  {q(\tTau, \tC \mid \Tau, C) \cancel{q(\tbx \mid \tTau, \tC)}}.
\end{multline*}
Here we use $p_b$ as an informative proposal distribution for $\tbx$ to achieve a better mixing Markov chain Monte Carlo simulation with reasonable acceptance probabilities. Without the use of an informative proposal, the proposed new tree is almost always rejected because the multinomial likelihood with the large sample size is very peaked.
Under the modified prior $p_b(\cdot)$, the resulting conditional posterior on $\bx$ remains entirely unchanged, $p_b(\bx \mid \Tau, C, \bn) = p(\bx \mid \Tau, C, \bn)$ [\cite{zhou2017pairclone}].

The described uniform tree proposal is in contrast to usual search
algorithms for trees that generate proposals from neighboring trees.
Such algorithms have the important drawback
that they quickly gravitate towards a local mode and then get stuck. A
possible approach to addressing this problem is to repeatedly restart
the algorithm from different starting trees. See
\cite{chipman1998bayesian} for more details. Our uniform tree proposal
combined with the data splitting scheme is another way to mitigate
this challenge, efficiently searching the tree space while keeping a
reasonable acceptance probability.

All posterior inference is contained in the posterior
distribution for $\bx, C$ and $\Tau$.
For example, the marginal posterior distribution of $C$ and $\Tau$
gives updates posterior probabilities for all possible values of $C$
and $\Tau$. But it is still useful to report point estimates. We use the
posterior modes $(\Chat, \Tauhat)$ as point estimates for $(C, \Tau)$,
and conditional on $\Chat$ and $\Tauhat$, 
we use the maximum a posteriori (MAP) estimator as an estimation for
the other parameters. 
The MAP is approximated as the MCMC sample with highest posterior
probability.
Let $\{ \bx^{(l)}, l = 1, \ldots, L \}$ be a set
of MCMC samples of $\bx$, and 
\begin{align*}
\hat{l} = \argmax_{l \in \{1, \ldots, L\}} \; p(\bn \mid \bx^{(l)}, \Tauhat, \Chat) \, p(\bx^{(l)} \mid \Tauhat, \Chat).
\end{align*}
We report point estimates as $\Zhat = \bZ^{(\hat{l})}$, $\what =
\bw^{(\hat{l})}$ and $\hat{\brho} = \brho^{(\hat{l})}$.

\section{Simulation Studies}
\label{sec:sim}
We  carried out  three simulation studies to assess the
performance of TreeClone.
We simulate multiple datasets with different number of
subclones ($C$), number of samples ($T$), average read depths ($\Nbar_{tk}$)
and left and right missing rates ($v_{tk2},v_{tk3}$) to test the
performance of the proposed model in different scenarios.  
In all simulation studies, we generate hypothetical read count data
for dozens of mutation pairs ($K = 100$ for simulations 1 and 2, and $K = 69$ for simulation 3), which is a typical number of mutation
pairs in a real tumor sample. 

\subsection{Simulation 1, Convergence Diagnostic and Sensitivity Analysis}
\paragraph{Simulation 1(a)}
We first validate TreeClone on 9 simulated datasets, one for each
combination of $C \in \{3,4,5\}$ and
average read depth $\bar{N}_{tk} \in \{50, 200, 1000\}$. For all 9 datasets, we
consider $T = 5$ samples and $K = 100$ mutation pairs. We randomly
generate the phylogenetic tree $\Tau$ and the genotype matrix
$\bZ$  from the prior model.  The subclone
proportions are simulated from
$\bw_t \sim \Dir(0.01, \sigma(15, 10, 5))$, 
$\Dir(0.01, \sigma(15, 10, 8, 5))$, 
$\Dir(0.01, \sigma(15, 10, 8, 5, 3))$
for $C = 3$, $4$ and $5$, respectively. Here $\sigma(x_1, \ldots, x_n)$ stands for a random permutation of $x_1, \ldots, x_n$. 
The noise fractions $\brho$ are generated from the prior with $d_1 = 1$. 
We mimic a typical rate of left (or right) missing reads by setting
$v_{tk2} = v_{tk3} = 0.3$, for all $t$ and $k$. The read depth
$N_{tk}$ is generated from a negative-binomial distribution, $N_{tk}
\sim \text{NB}(r_N, p_N)$, to reflect the over-dispersion of read
depth in sequencing data.
We fix $r_N$ and $p_N$ such that $\E(N_{tk}) =
\bar{N}_{tk}$ and $\sd(N_{tk}) = \bar{N}_{tk} / 5$ for
$\bar{N}_{tk} = 50$, $200$ and $1000$. Finally, the read counts 
$\{n_{tkg}\}$ are simulated from the multinomial sampling model
\eqref{eq:multi-sampling}.

We fit the model with the following hyperparameters: $\alpha = 0.5$,
$\beta = 0.5$, $d = 0.5$, $d_0 = 0.03$, $d_1 = 1$, where the values of
$\alpha$ and $\beta$ imply mild penalty for deep and bushy trees
[\cite{chipman1998bayesian}], and other hyperparameters are default
non-informative choices.  We set $a_p = d, b_p = d_0 + (C-1)d$ for
given $C$ as a non-informative prior choice and set $\lambda = 2K/C$
to express our prior belief that about half of the mutations occur
uniformly at each generation. We use $C_{\min} = 2$ and $C_{\max} = 5$
as the range of $C$ for computational efficiency. In principle
$C_{\max}$ can be any finite number. However, the size of the tree
space grows exponentially with $C_{\max}$, so that the time needed for
achieving a good mixing Markov chain grows exponentially.
We set the training set fraction (in the trans-dimensional MCMC
implementation) to $b = 0.95$, which we found to
performs well in all simulation studies.  We run a total of 8000 MCMC
iterations. Discarding the first 3000 draws as initial burn-in, we
have a Monte Carlo sample with 5000 posterior draws.

We evaluate the performance of TreeClone in estimating
the number of subclones $C$, phylogeny $\Tau$,  genotype $\bZ$ and
subcone proportions $\bw$.
To this end, we define the reconstruction error rates
$C_\err = I(\Chat \neq C)$, $\Tau_\err = I(\Tauhat \neq \Tau)$,
\begin{align*}
Z_{\text{err}} = \frac{1}{K(C-1)} \min_{\sigma} \left( \sum_{k,c} I(\hat{\bz}_{k\sigma(c)} \neq \bz_{kc})\right) 
\end{align*}
and $w_{\text{err}} = \sum_{t, c} | \hat{w}_{t\sigma(c)} - w_{tc} | /
(TC)$, similar to \cite{marass2017phylogenetic}.
Here $\sigma$ is a permutation of columns of $\bZ$ to account for
label-switching of subclones, and the same permutation is imposed on
columns of $\bw$. The simulation results are summarized in Table
\ref{tbl:sim1_a}. For all 9 simulated datasets TreeClone nicely
recovers the truth and attains reasonably low reconstruction errors.

\begin{table}[h!]
\begin{center}
\begin{tabular}{|c|c|c|c|c|}
\hline
\backslashbox{$\bar{N}_{tk}$}{$C$} & 3 & 4 & 5 \\\hline
50 & 
0, 0, 0.00, 0.09 & 
0, 0, 0.03, 0.07 & 
0, 0, 0.07, 0.06  \\\hline
200 & 
0, 0, 0.00, 0.15 & 
0, 0, 0.00, 0.09 &  
0, 0, 0.00, 0.05  \\\hline
1000 & 
0, 0, 0.00, 0.14 & 
0, 0, 0.00, 0.10 & 
0, 0, 0.00, 0.08 \\\hline
\end{tabular}
\end{center}
\caption{Simulation 1(a). Summary of simulation results for 9
   combinations of $C$ and $\bar{N}_{tk}$. 
  Each cell of the table reports $(C_\err,
  \Tau_\err, Z_{\text{err}}, w_{\text{err}})$ as an average over three
  runs of TreeClone with different random number seeds.}
\label{tbl:sim1_a}
\end{table}

To assess the convergence of the MCMC algorithm, for each simulated
dataset we run the algorithm three times with different random
seeds. 
We use the log-posterior values 
to calculate a potential scale reduction factor (PSRF)
[\cite{gelman1992inference}] for the three Markov chains.
A PSRF close to 1 indicates convergence of the Markov chain to the
target distribution.
The results are reported in Table \ref{tbl:sim1_a_diag}.
Next, to assess the identifiability of the parameters in the TreeClone
model, we calculate frequentist coverage probabilities of
$95\%$ posterior  credible intervals for $p_{tkg}$.
The results are shown as the second entry in each cell of Table
\ref{tbl:sim1_a_diag}.

\begin{table}[h!]
\begin{center}
\begin{tabular}{|c|c|c|c|c|}
\hline
\backslashbox{$\bar{N}_{tk}$}{$C$} & 3 & 4 & 5 \\\hline
50 & 
1.06, $100\%$ & 
1.08, $85\%$ & 
1.33, $96\%$  \\\hline
200 & 
1.13, $92\%$ & 
1.04, $86\%$ &  
1.38, $88\%$  \\\hline
1000 & 
1.05, $100\%$ & 
1.11, $83\%$ & 
3.72, $94\%$ \\\hline
\end{tabular}
\end{center}
\caption{Simulation 1(a). Convergence diagnostic and frequentist
  coverage probabilities for 9 simulated datasets. For each simulated
  dataset, we run TreeClone three times with different random seeds. Each
  cell of the table reports the PSRF (for log-posterior values) and average
  coverage rate (of $95\%$ credible intervals for $p_{tkg}$), averaged
  over the three chains.}
\label{tbl:sim1_a_diag}
\end{table}

\paragraph{Simulation 1(b)}
Next, we  vary average read depth $\bar{N}_{tk}$ and $T$. 
Again we simulate 9 datasets,
one for each combination of $T$ and $\Nbar_{tk}$ with $T \in \{1, 3,
5\}$ and $\bar{N}_{tk}  \in \{50, 200, 1000\}$. 
We assume the same genotype matrix $\bZ$ for all the 9 datasets with
$C = 4$ subclones and $K = 100$ mutation pairs. The parameters are
generated from the prior model, and $N_{tk}$ is generated from
$\text{NB}(r_N, p_N)$, as in simulation 1(a).

\begin{table}[h!]
\begin{center}
\begin{tabular}{|c|c|c|c|c|}
\hline
\backslashbox{$\bar{N}_{tk}$}{$T$} & 1 & 3 & 5 \\\hline
50 & 
0, 0, 0.29, 0.05 &  
0, 0, 0.11, 0.09 &  
0, 0, 0.03, 0.08  \\\hline
200 & 
0, 0, 0.16, 0.04 & 
0, 0, 0.01, 0.07 &  
0, 0, 0.00, 0.08  \\\hline
1000 & 
0, 0, 0.13, 0.02 & 
0, 0, 0.00, 0.12 & 
0, 0, 0.00, 0.10  \\\hline
\end{tabular}
\end{center}
\caption{Simulation 1(b). Summary of posterior inference
  for 9 combinations of $T$ and $\Nbar_{tk}$.
  Each cell of the table reports $(C_\err, \Tau_\err,
  Z_{\text{err}}, w_{\text{err}})$.} 
\label{tbl:sim1_b}
\end{table}

The simulation results are summarized in Table \ref{tbl:sim1_b}. Even
with only one sample and low read depth, TreeClone
 can reliably estimate 
 $C$ and $\Tau$ (although it does not perfectly recover
$\bZ$ due to the limited amount of data).

\paragraph{Simulation 1(c)}
Finally, we assess  sensitivity with respect to
 the rates of left and right missing mutation pairs $v_{tk2}$ and
$v_{tk3}$. 
Missingness is simply due to the length of a short
read is not long enough to cover both loci in a mutation pair. Thus,
missingness is non-informative, and we assume $v_{tk2} = v_{tk3}$
in the simulation for simplicity. We simulate 5 datasets with missing
rates $v_{tk2} = v_{tk3} = 0$, $0.1$, $0.25$, $0.4$ and $0.5$. The
extreme case $v_{tk2} = v_{tk3} = 0.5$ implies that no
complete mutation pair reads are recorded.
For all the 5 datasets, we consider $T = 5$ samples and
average read depth $\bar{N}_{tk} = 200$, and we assume the same
genotype $\bZ$ with $C = 4$ subclones and $K = 100$ mutation
pairs. The parameters are generated from the prior model as in simulation
1(a) and (b). 
The simulation results are summarized in Table
\ref{tbl:sim1_c}. TreeClone maintains high reconstruction accuracy across
all scenarios. 

\begin{table}[h!]
\begin{center}
\begin{tabular}{|c|c|c|c|c|c|}
  \hline
  \multicolumn{5}{|c|}{$v_{tk2}=v_{tk3}$} \\\hline
$0$ & $0.1$ & $0.25$ & $0.4$ & $0.5$  \\\hline
0, 0, 0.00, 0.08 & 
0, 0, 0.00, 0.09 & 
0, 0, 0.00, 0.08 & 
0, 0, 0.00, 0.07 & 
0, 0, 0.12, 0.07  \\\hline
\end{tabular}
\end{center}
\caption{Simulation 1(c). Summary of posterior inference in
  simulated datasets under a range of values for $v_{tk2}=v_{tk3}$.
  Each entry reports $(C_\err, \Tau_\err, Z_{\text{err}},
  w_{\text{err}})$. } 
\label{tbl:sim1_c}
\end{table}

\paragraph{ Using only marginal SNVs}
We  consider another simulation assuming that we 
observe only marginal SNVs that are not phased.
We treat the marginal SNVs as right missing reads, i.e. $v_{tk2} = 0$
and $v_{tk3} = 1$. We simulate one dataset under this scenario. The
reconstruction errors are $(C_\err, \Tau_\err, Z_\err, w_\err) = (0,
0, 0.46, 0.1)$.
The number of subclones and phylogenetic tree are correctly
identified. The genotype reconstruction error $Z_\err$ is high because
the data do not provide any information for inferring the unobserved
locus $z_{kcj2}$ for all $k$, $c$ and $j$. For a fair comparison we
re-define $Z_{\text{err}}^{\text{SNV}} = \min_{\sigma} \left(
\sum_{k,c} I(\hat{z}_{k\sigma(c)j1} \neq z_{kcj1} ~\text{for $j = 1$
  or $2$})\right) / \left( K(C-1) \right)$. The re-defined
reconstruction error is $Z_{\text{err}}^{\text{SNV}} = 0.00$.

\subsection{Simulation 2 and comparison with alternatives}
In the second simulation study, we compare the performance of
TreeClone with existing methods.  We consider $T = 1$ sample, which is
the case for most real-world tumor cases (due to the challenge in
obtaining multiple samples from a patient).
In practice we notice that methods using only marginal SNV data find
it hard to identify branching in a phylogenetic tree. We use this
simulation to illustrate the information gain by using mutation pairs
data.

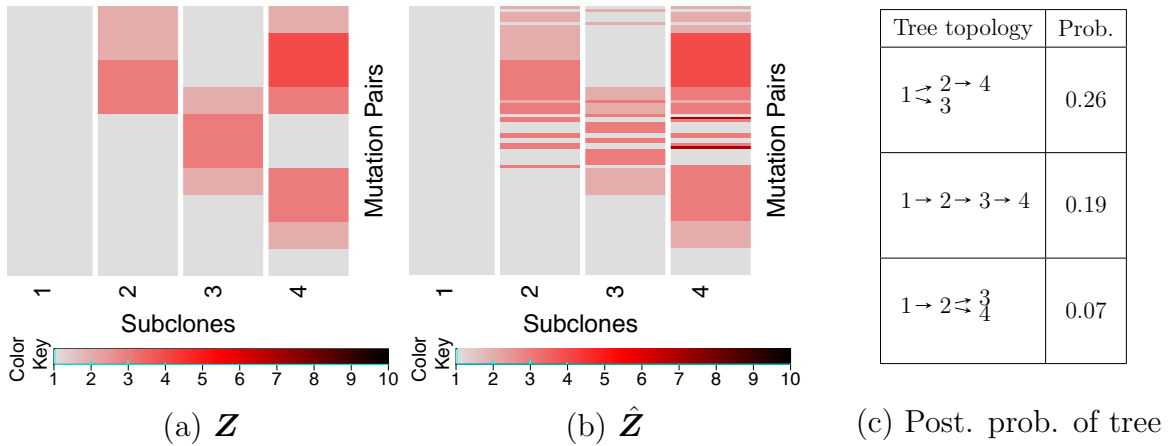
\begin{figure}[h!]
\begin{center}
\begin{subfigure}[t]{.325\textwidth}
\centering
\includegraphics[width=\textwidth]{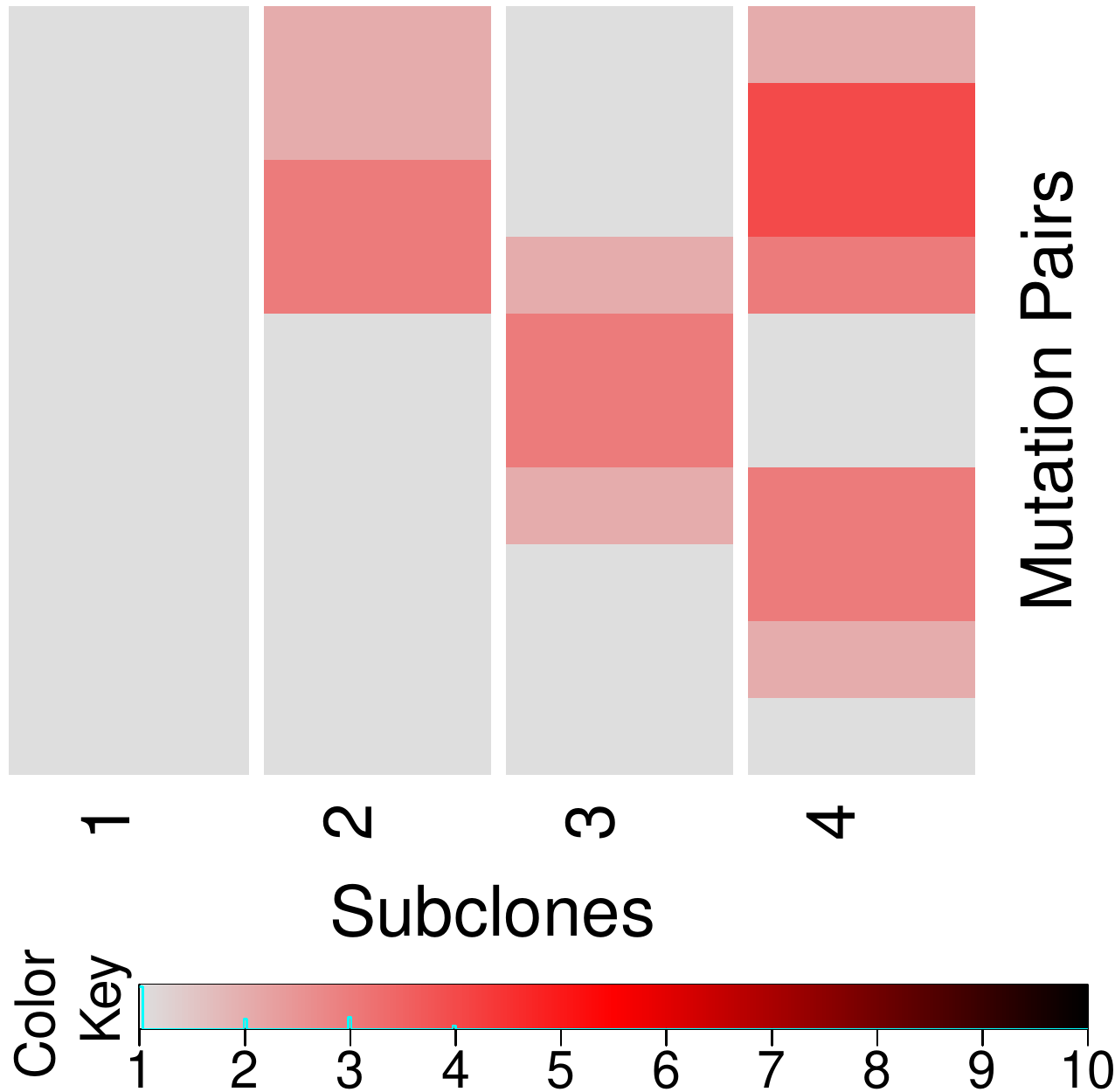}
\caption{$\bZ$}
\end{subfigure}
\begin{subfigure}[t]{.325\textwidth}
\centering
\includegraphics[width=\textwidth]{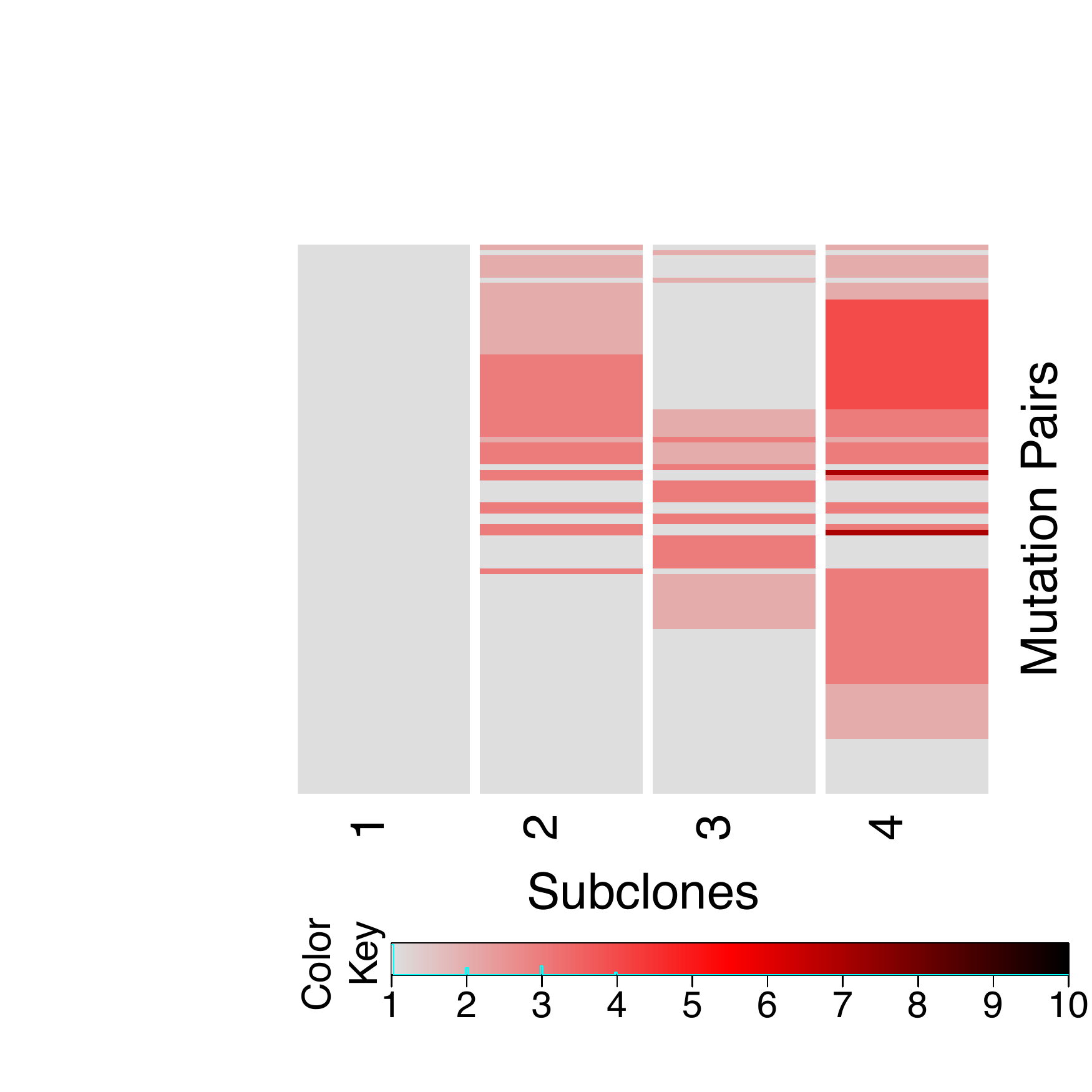}
\caption{$\Zhat$}
\end{subfigure}
\begin{subfigure}[t]{.325\textwidth}
\centering
\vspace{-50mm}
\scalebox{0.74}{
\begin{tabular}{|>{\raggedright\arraybackslash}m{25mm}|>{\centering\arraybackslash}m{10mm}|@{}m{0pt}@{}}
\hline
\multicolumn{1}{|c|}{Tree topology} & Prob. & \\ \hline
\begin{tikzpicture}[grow=right, sloped]
\tikzstyle{level 1}=[level distance=0.7cm, sibling distance=0.4cm]
\tikzstyle{level 2}=[level distance=0.7cm, sibling distance=0.3cm]
\node[bag] {1}
    child {
        node[bag] {3}
        edge from parent [-stealth]
    }
    child {
        node[bag] {2}        
        child {
                node[bag] {4}
                edge from parent [-stealth]
            }
        edge from parent [-stealth]       
    };
\end{tikzpicture} & 0.26 &  \\[3em]
\hline
\begin{tikzpicture}[grow=right, sloped]
\tikzstyle{level 1}=[level distance=0.7cm, sibling distance=0.4cm]
\tikzstyle{level 2}=[level distance=0.7cm, sibling distance=0.3cm]
\node[bag] {1}
    child {
        node[bag] {2}        
        child {
                node[bag] {3}
                child {
                  node[bag] {4}
                  edge from parent [-stealth]
                }
                edge from parent [-stealth]
            }
        edge from parent [-stealth]       
    };
\end{tikzpicture}
 & 0.19  & \\[3em]
 \hline
\begin{tikzpicture}[grow=right, sloped]
\tikzstyle{level 1}=[level distance=0.7cm, sibling distance=0.4cm]
\tikzstyle{level 2}=[level distance=0.7cm, sibling distance=0.3cm]
\node[bag] {1}
    child {
        node[bag] {2}        
        child {
                node[bag] {4}
                edge from parent [-stealth]
            }
        child {
                node[bag] {3}
                edge from parent [-stealth]
            }    
        edge from parent [-stealth]       
    };
\end{tikzpicture}
 & 0.07 &  \\[3em]
 \hline
\end{tabular}
}
\vspace{2.5mm}
\caption{Post. prob. of tree}
\end{subfigure}
\end{center}
\caption{Simulation 2. Simulation truth $\bZ$ (a) and posterior inference under TreeClone (b, c). }
\label{fig:sim1}
\end{figure}

We consider $K = 100$ mutation pairs and assume a simulation truth
with $C = 4$ latent subclones with a true phylogenetic tree as
\begin{center}
\begin{tikzpicture}[grow=right, sloped]
\node[bag] {1}
    child {
        node[bag] {3.}
        edge from parent [-stealth]
    }
    child {
        node[bag] {2}        
        child {
                node[bag] {4}
                edge from parent [-stealth]
            }
        edge from parent [-stealth]       
    };
\end{tikzpicture}
\end{center}
Fig.~\ref{fig:sim1}(a) shows the true
underlying subclonal genotypes. 
We use a heatmap to show the subclone matrix $\bZ$, where colors from
light gray to red to black are used to represent genotypes $\bz^{(1)}$
to $\bz^{(10)}$.  
The subclone weights are simulated from $\Dir(0.01, \sigma(15, 10, 8, 5))$. 
For the single sample in this simulation we get $\bw = (0.000$,
$0.135$, $0.169$, $0.470$, $0.226)$. 
We generate $\brho$ from the prior model with $d_1 = 1$, and we set
$v_{tk2} = v_{tk3} = 0.3$, for all $t$ and $k$.  
We generate the read depth $N_{tk} \sim \text{NB}(r_N, p_N)$ with
$\E(N_{tk}) = 500$ and $\sd(N_{tk}) = 100$. 

The hyperparameters are set as in Simulation 1. To explore a larger
tree space, we set $C_{\max} = 6$, run a total of 13000 MCMC
iterations and discard the first 3000 draws as initial burn-in.

Posterior inference is summarized in
Fig.~\ref{fig:sim1}(b, c). Fig.~\ref{fig:sim1}(c) shows 
the top three tree topologies and corresponding posterior
probabilities.
The posterior mode
recovers the true phylogeny. Fig.~\ref{fig:sim1}(b) shows the
estimated genotypes conditional on the posterior modes $(\Chat, \Tauhat)$. 
Some mismatches are due to
the single sample and limited read depth. The estimated subclone
proportions are 
$\what = (0.000$, $0.103$, $0.162$, $0.498$, $0.237)$,
which agrees with the truth.

\begin{figure}[h!]
\begin{center}
\begin{subfigure}[t]{.24\textwidth}
\centering
\includegraphics[width=\textwidth]{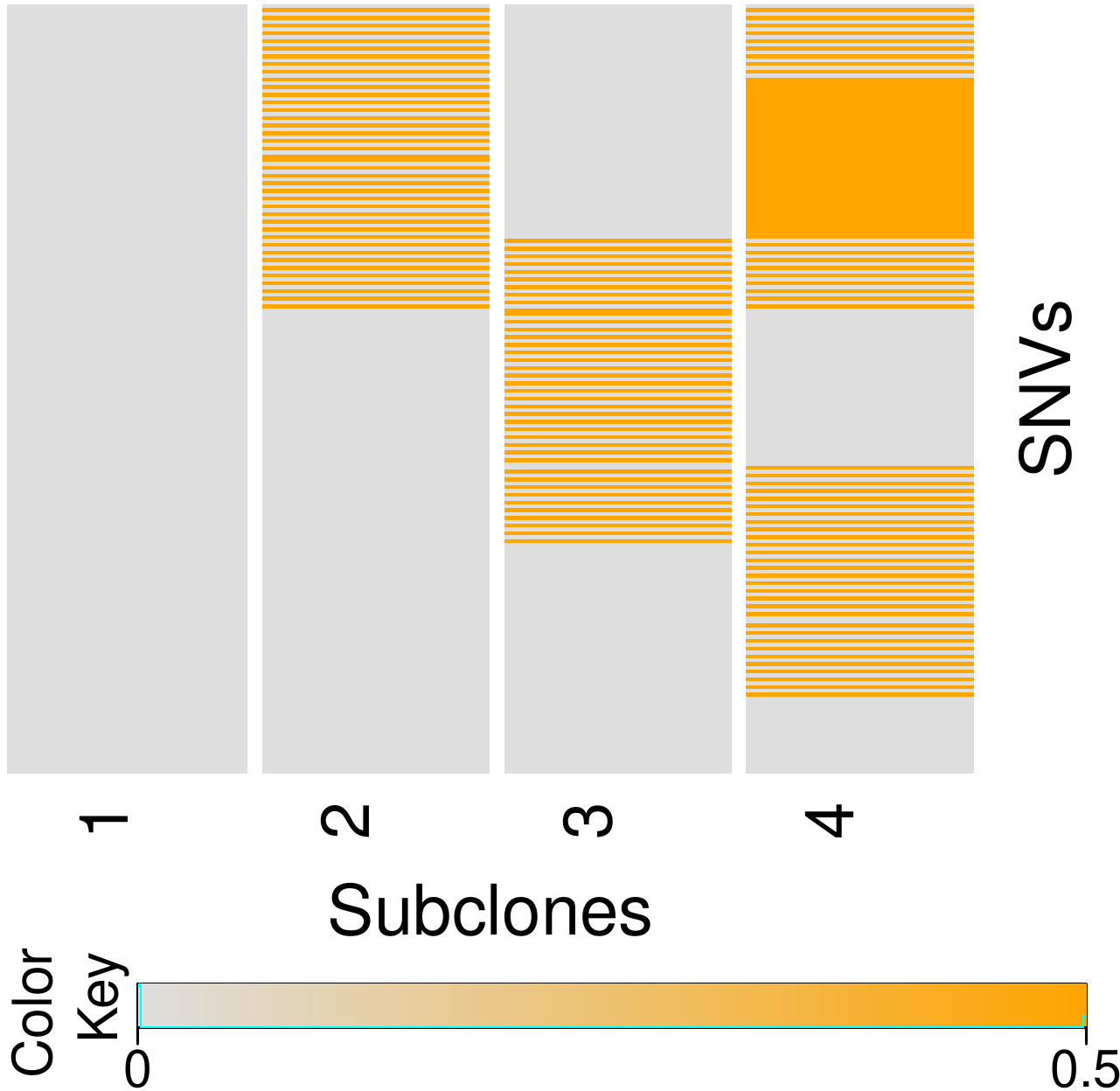}
\caption{$\bZ^{\text{Cloe}}$}
\label{fig:sim1_Z_Cloe}
\end{subfigure}
\begin{subfigure}[t]{.24\textwidth}
\centering
\includegraphics[width=\textwidth]{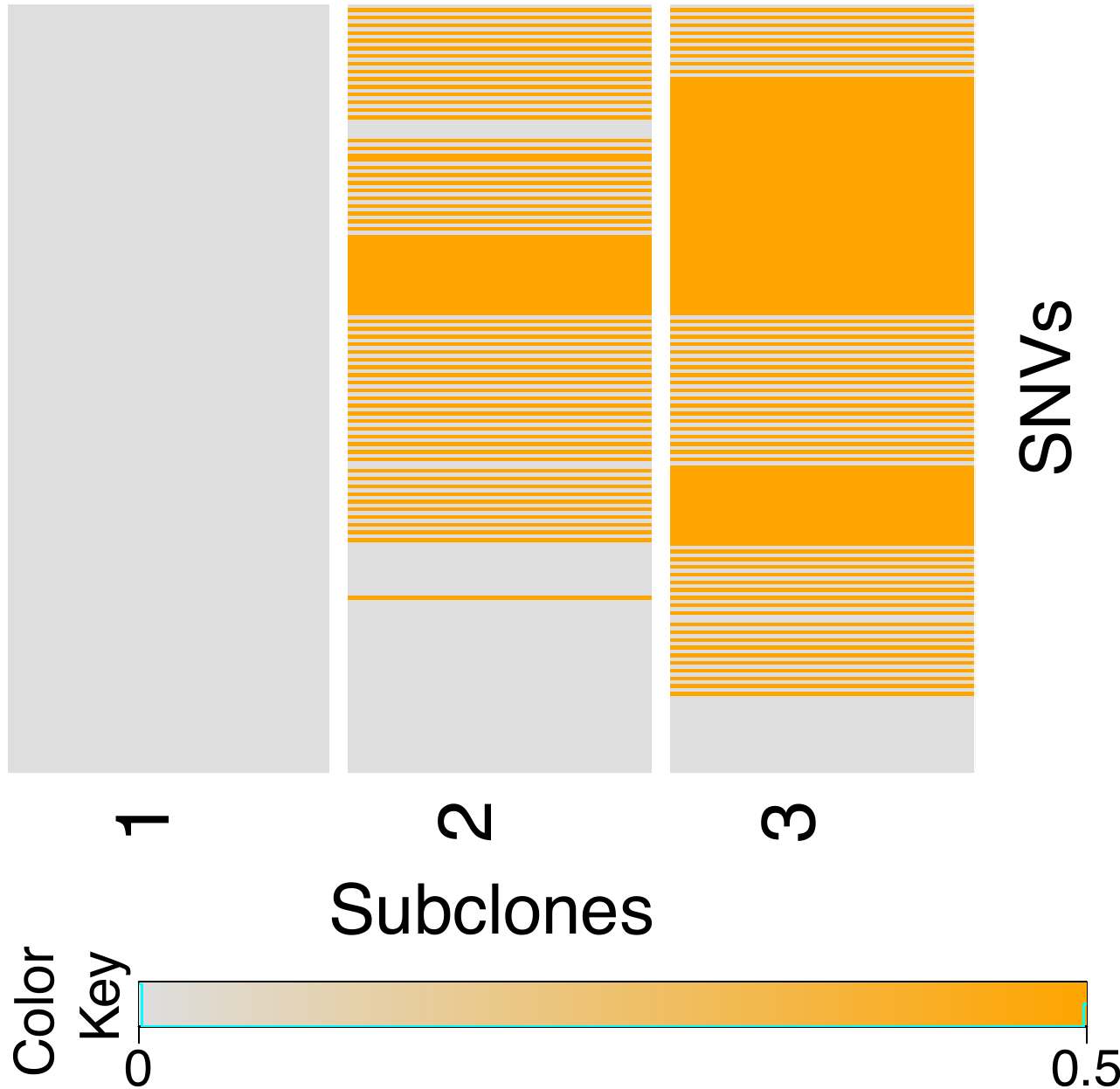}
\caption{$\Zhat^{\text{Cloe}}$}
\label{fig:sim1_Zhat_Cloe}
\end{subfigure}
\begin{subfigure}[t]{.24\textwidth}
\centering
\includegraphics[width=\textwidth]{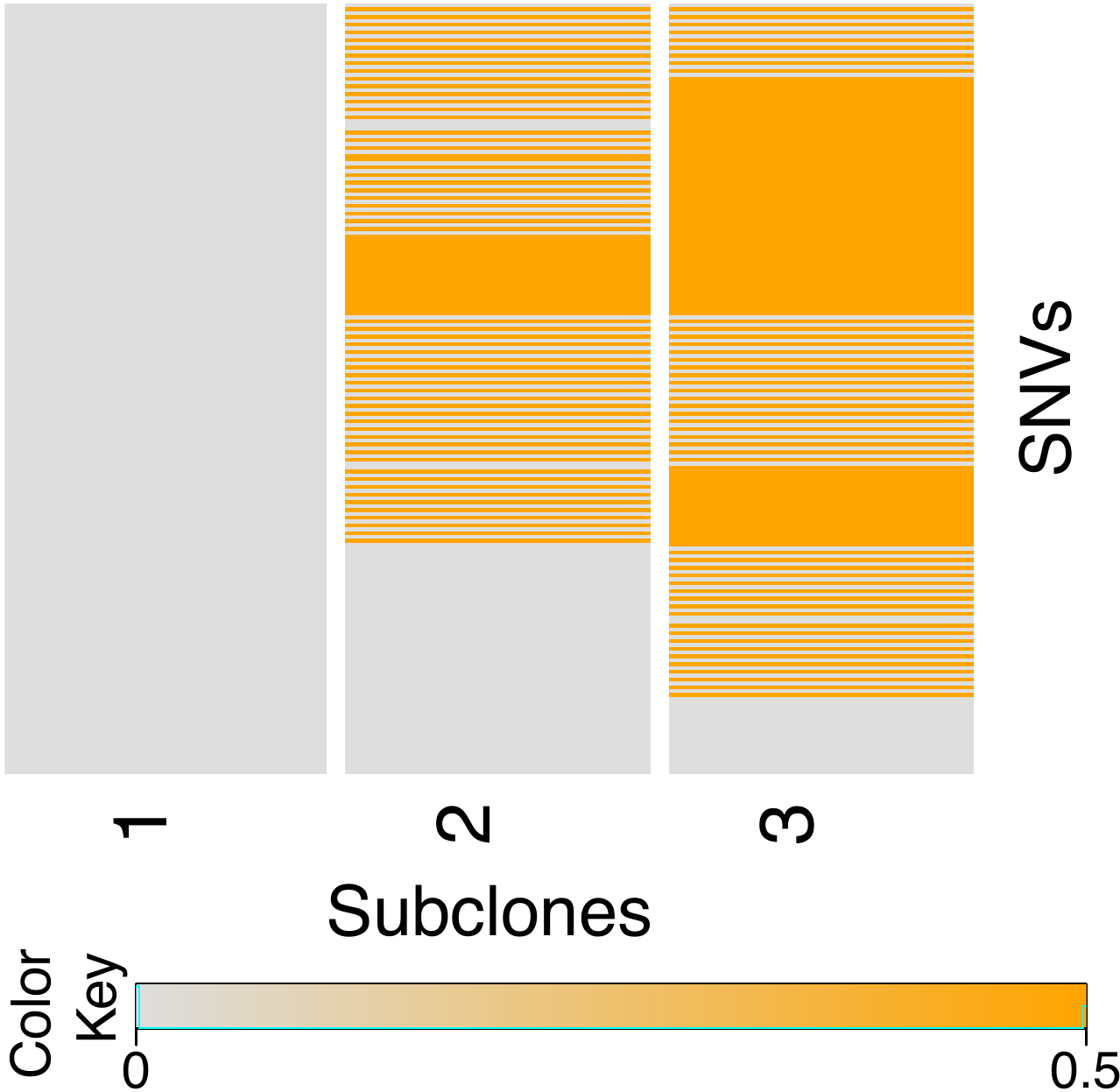}
\caption{$\Zhat^{\text{PWGS}}$}
\label{fig:sim1_Zhat_PWGS}
\end{subfigure}
\begin{subfigure}[t]{.24\textwidth}
\centering
\includegraphics[width=\textwidth]{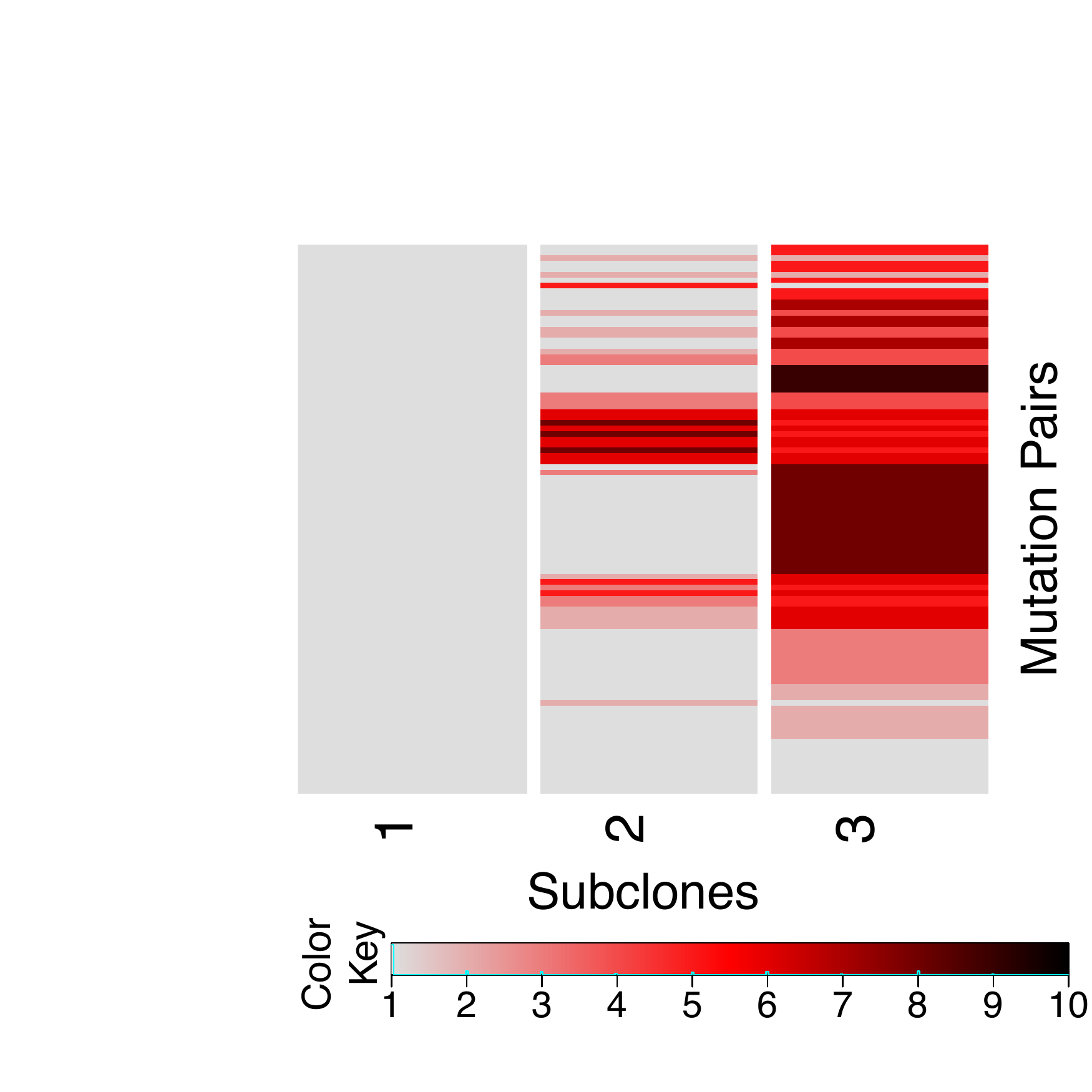}
\caption{$\Zhat^{\text{PairClone}}$}
\label{fig:sim1_Zhat_PairClone}
\end{subfigure}
\end{center}
  \caption{Simulation 2. Simulation truth $\bZ^{\text{Cloe}}$ (a), and posterior inference under Cloe (b), PhyloWGS (c) and PairClone (d). }
  \label{fig:sim1_compare}
\end{figure}

\paragraph{Comparison with alternatives}
There is no other subclone calling method based on
paired-end read data that also infers phylogeny. 
We therefore compare with other similar model-based
approaches. In particular, we use Cloe [\cite{marass2017phylogenetic}], 
PhyloWGS [\cite{jiao2014inferring, deshwar2015phylowgs}] 
and PairClone [\cite{zhou2017pairclone}]
for inference with the same simulated data. 
Cloe and PhyloWGS infer phylogeny but take marginal SNV data as input.
 For these methods we  therefore discard the phasing information in
mutation pairs and only record marginal mutation counts for SNVs.
The simulation truth for Cloe and PhyloWGS is shown in
Fig. \ref{fig:sim1_Z_Cloe}. 
The orange color means a heterozygous mutation at the corresponding
SNV locus.  PairClone takes the same mutant read counts and read
depths for mutation pairs as input but
 uses a different probability model that does not allow to  
infer the phylogenetic tree.

\textbf{Cloe} infers clonal genotypes and phylogeny based on a similar feature allocation model.
We run Cloe with the default hyperparameters for the same number of
13000 iterations with the first 3000 draws as initial burn-in.
Based on the MAP estimate  (over $2 \leq C \leq 6$),  Cloe
reports $C=3$ subclones with phylogeny
$1 \rightarrow 2 \rightarrow 3$, and the subclone genotypes are shown in
Fig. \ref{fig:sim1_Zhat_Cloe} with subclone proportions
$\what^{\text{Cloe}} = (0.555$, $0.223$, $0.222)$.


\textbf{PhyloWGS}, on the other hand, infers clusters of mutations and phylogeny. One can then make phylogenetic analysis to conjecture subclones and genotypes.
Let $\tilde{\phi}_i$ denote the fraction of cells with a variant allele at
locus $i$. The $\tilde{\phi}_i$'s are latent quantities related to the observed VAF for each SNV. PhyloWGS infers the phylogeny by clustering SNVs with
matching $\tilde{\phi}_i$'s under a tree-structured prior for the unique values $\phi_j$. In particular, they use the tree-structured stick breaking process (TSSB)
[\cite{adams2010tree}]. 
The TSSB implicitly defines a prior
on the formation of subclones, including a prior on $C$ and the number of
novel loci that arise in each subclone
(in contrast, TreeClone explicitly defines these model features,
allowing easier prior control on $C$ and $\mathcal{M}_c$).
We run PhyloWGS with the default hyperparameters and 2500 iterations
with a burn-in of 1000 samples. We only consider loci with VAF $> 0$
as for PhyloWGS the other loci do not provide information on clustering.
We then report cluster sizes and phylogeny based on MAP
estimate. PhyloWGS reports 3 subclones with phylogeny $0 \rightarrow 1
(79, 0.438) \rightarrow 2 (53, 0.220)$, where 0 refers to the normal
subclone, and the numbers in the brackets refer to the cluster sizes
and cellular prevalences. The conjectured subclone genotypes are shown
in Fig. \ref{fig:sim1_Zhat_PWGS}, with subclone proportions
$\what^{\text{PWGS}} = (0.562$, $0.218$, $0.220)$.

Inferences under Cloe and PhyloWGS do not entirely recover the
truth.
 Let $\mathcal{M}_c$ denote the new mutations that are gained by
subclone $c$. 
The reason for the failure to recover the simulation truth  is
probably that the common mutations of subclones 2 
and 4 ($\mathcal{M}_2$ with a cellular prevalence of $0.169+0.226$)
have a similar cellular prevalence as the mutations of subclone 3
($\mathcal{M}_3$ with a cellular prevalence of $0.470$).
Therefore, Cloe infers that $\mathcal{M}_2$
and $\mathcal{M}_3$ belong to the same subclone
($\mathcal{M}_2^{\text{Cloe}} \approx \mathcal{M}_2 \cup
\mathcal{M}_3$ and $\mathcal{M}_3^{\text{Cloe}} \approx
\mathcal{M}_4$). Similarly, PhyloWGS clusters $\mathcal{M}_2$ and
$\mathcal{M}_3$ together.
Using more informative mutation pairs data, TreeClone  can
infer  that $\mathcal{M}_2$ and $\mathcal{M}_3$ belong to different
subclones. 
The inclusion of phasing information from the paired-end read data
increases statistical power in recovering the underlying structure.


\textbf{PairClone} uses the same mutation pairs data and same sampling model to infer clonal genotypes. However, PairClone uses a finite categorical Indian buffet process prior for $\bZ$, which assumes independence among the subclones and does not infer phylogeny. PairClone infers 3 subclones with genotypes shown in Fig. \ref{fig:sim1_Zhat_PairClone}. The estimated subclone proportions are $\what^{\text{PairClone}} = (0.594$, $0.229$, $0.177)$, similar to Cloe and PhyloWGS's results.
PairClone does not entirely recover the truth, probably because not
imposing the tree structure reduces the identifiability of the
problem.

\paragraph{Comparison using additional marginal SNVs}
An NGS dataset contains many more marginal SNVs than phased mutation
pairs. These additional marginal SNVs can be utilized by 
methods such as Cloe and PhyloWGS.
For an illustration of the information gain by using more SNVs, we run
Cloe and PhyloWGS with a larger set of SNVs that contains
 400 SNVs in addition to the 200 SNVs that are part of the 100
mutation pairs, i.e., a total of 600 SNVs. 
We assume a genotype matrix $\bZ_{+} = (\bZ, \bZ, \bZ)'$,
i.e. repeating the rows of $\bZ$ three times, and keep the other
parameters unchanged from before.

Using 600 SNVs, Cloe infers two subclones with phylogeny $1
\rightarrow 2$ and proportions $\what_+^{\text{Cloe}} = (0.641$,
$0.359)$. PhyloWGS infers three (conjectured) subclones with phylogeny
$1 \rightarrow 2 \rightarrow 3$ and  proportions
$\what_+^{\text{PWGS}} = (0.562$, $0.212$, $0.226)$. The results
suggest that the additional 400 SNVs do not  contribute much
information about the branching in the phylogenetic tree. 


\subsection{Simulation 3}
In the third simulation, we evaluate the performance of the proposed
approach on multiple samples. We generate the simulated data by
mimicking a real-world lung cancer data (see Section~\ref{sec:real}). Following that dataset, we
consider $T = 4$ tissue samples and $K = 69$ mutation pairs.  The simulation truth $\bZ$ and
$\bw$ are generated by fitting TreeClone on the lung cancer
dataset. We get $C = 6$ subclones with phylogeny

\begin{center}
\begin{tikzpicture}[grow=right, sloped]
\node[bag] {1}
    child {
        node[bag] {5}
            child {
                node[bag] {6.}
                edge from parent [-stealth]
            }
            edge from parent [-stealth]
    }
    child {
        node[bag] {2}        
        child {
                node[bag] {4}
                edge from parent [-stealth]
            }
            child {
                node[bag] {3}
                edge from parent [-stealth]
            }
        edge from parent [-stealth]       
    };
\end{tikzpicture}
\end{center}

\begin{figure}[h!]
\begin{center}
\begin{subfigure}[t]{.325\textwidth}
\centering
\includegraphics[width=\textwidth]{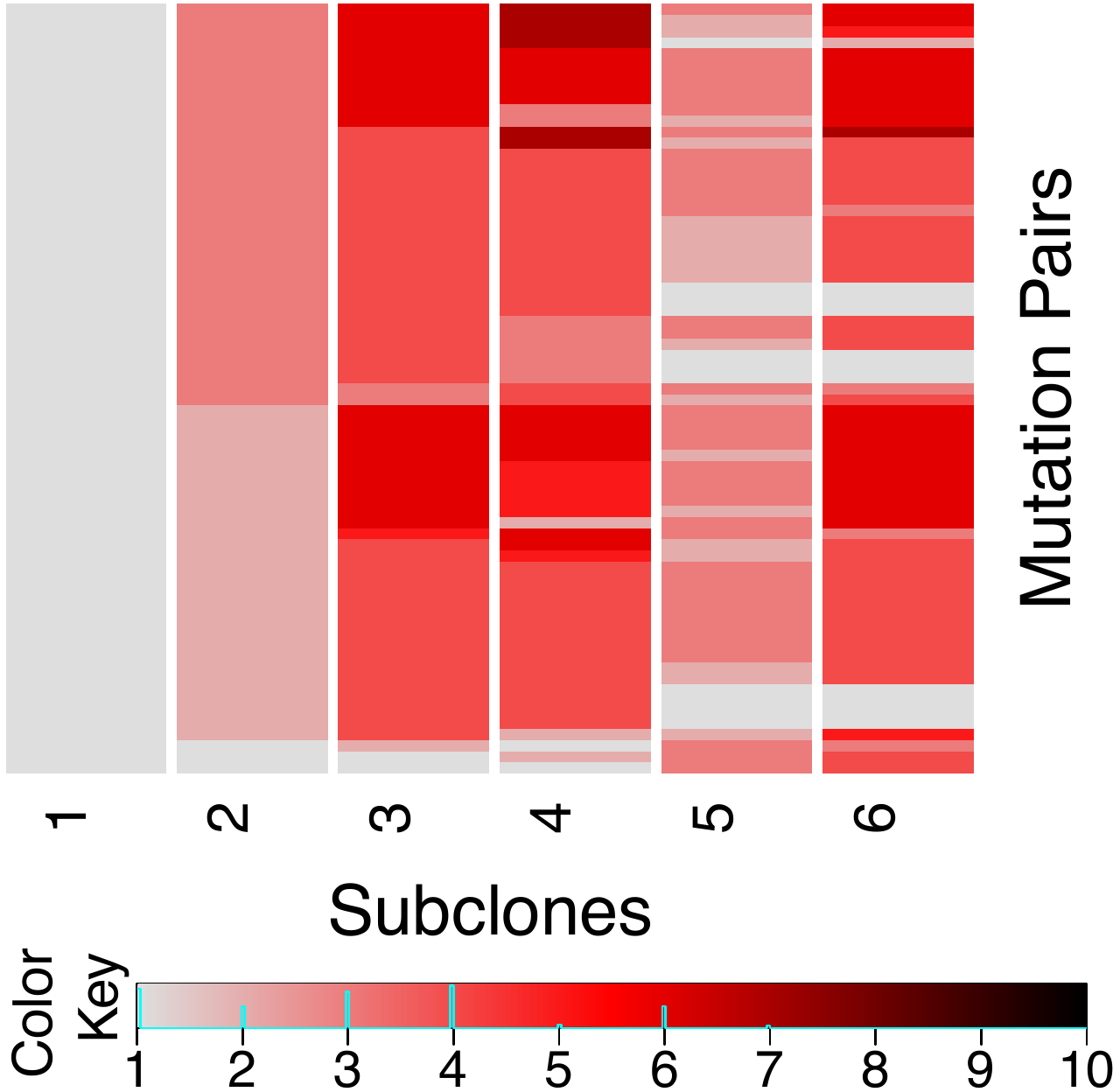}
\caption{$\bZ$}
\end{subfigure}
\begin{subfigure}[t]{.325\textwidth}
\centering
\includegraphics[width=\textwidth]{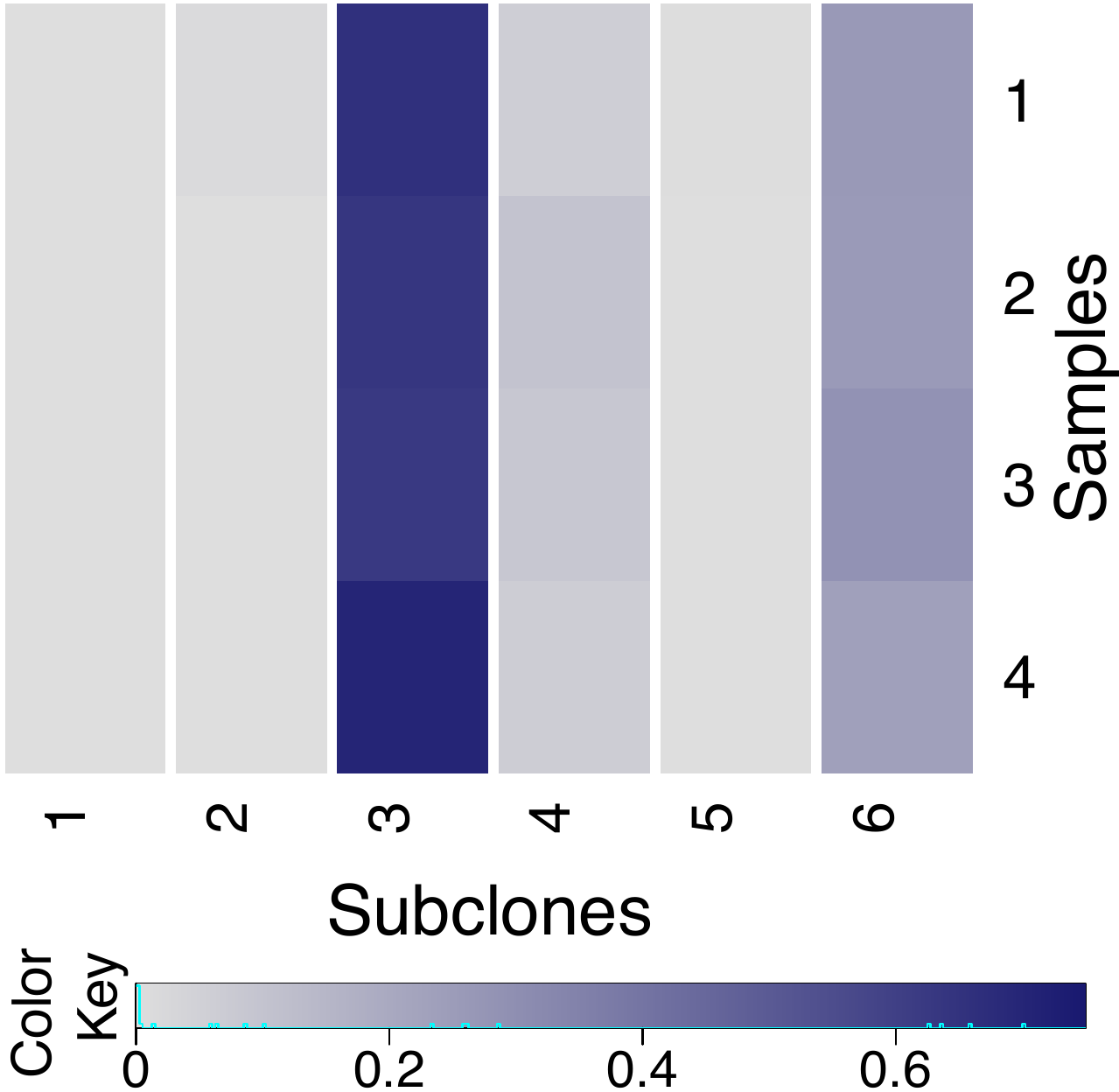}
\caption{$\bw$}
\end{subfigure}
\begin{subfigure}[t]{.325\textwidth}
\centering
\vspace{-50mm}
\scalebox{0.74}{
\begin{tabular}{|>{\raggedright\arraybackslash}m{25mm}|>{\centering\arraybackslash}m{10mm}|@{}m{0pt}@{}}
\hline
\multicolumn{1}{|c|}{Tree topology} & Prob. & \\ \hline
\begin{tikzpicture}[grow=right, sloped]
\tikzstyle{level 1}=[level distance=0.7cm, sibling distance=0.4cm]
\tikzstyle{level 2}=[level distance=0.7cm, sibling distance=0.3cm]
\node[bag] {1}
    child {
        node[bag] {5}
            child {
                node[bag] {6}
                edge from parent [-stealth]
            }
            edge from parent [-stealth]
    }
    child {
        node[bag] {2}        
        child {
                node[bag] {4}
                edge from parent [-stealth]
            }
            child {
                node[bag] {3}
                edge from parent [-stealth]
            }
        edge from parent [-stealth]       
    };
\end{tikzpicture} & 0.38 &  \\[3em]
\hline
\begin{tikzpicture}[grow=right, sloped]
\tikzstyle{level 1}=[level distance=0.7cm, sibling distance=0.6cm]
\tikzstyle{level 2}=[level distance=0.7cm, sibling distance=0.3cm]
\node[bag] {1}
    child {
        node[bag] {6}
            child {
                node[bag] {7}
                edge from parent [-stealth]
            }
            edge from parent [-stealth]
    }
    child {
        node[bag] {2} 
        child {
                node[bag] {5}
                edge from parent [-stealth]
            }       
        child {
                node[bag] {4}
                edge from parent [-stealth]
            }
            child {
                node[bag] {3}
                edge from parent [-stealth]
            }
        edge from parent [-stealth]       
    };
\end{tikzpicture}
 & 0.11  & \\[3em]
 \hline
\begin{tikzpicture}[grow=right, sloped]
\tikzstyle{level 1}=[level distance=0.7cm, sibling distance=0.6cm]
\tikzstyle{level 2}=[level distance=0.7cm, sibling distance=0.3cm]
\node[bag] {1}
    child {
        node[bag] {6}
            child {
                node[bag] {7}
                edge from parent [-stealth]
            }
            edge from parent [-stealth]
    }
    child {
        node[bag] {2} 
        child {
                node[bag] {5}
                edge from parent [-stealth]
            }       
            child {
                node[bag] {3}
                child {
                  node[bag] {4}
                  edge from parent [-stealth]
                }
                edge from parent [-stealth]
            }
        edge from parent [-stealth]       
    };
\end{tikzpicture}
 & 0.09 &  \\[3em]
 \hline
\end{tabular}
}
\vspace{2.5mm}
\caption{Post. prob. of tree}
\end{subfigure}
\begin{subfigure}[t]{.325\textwidth}
\centering
\includegraphics[width=\textwidth]{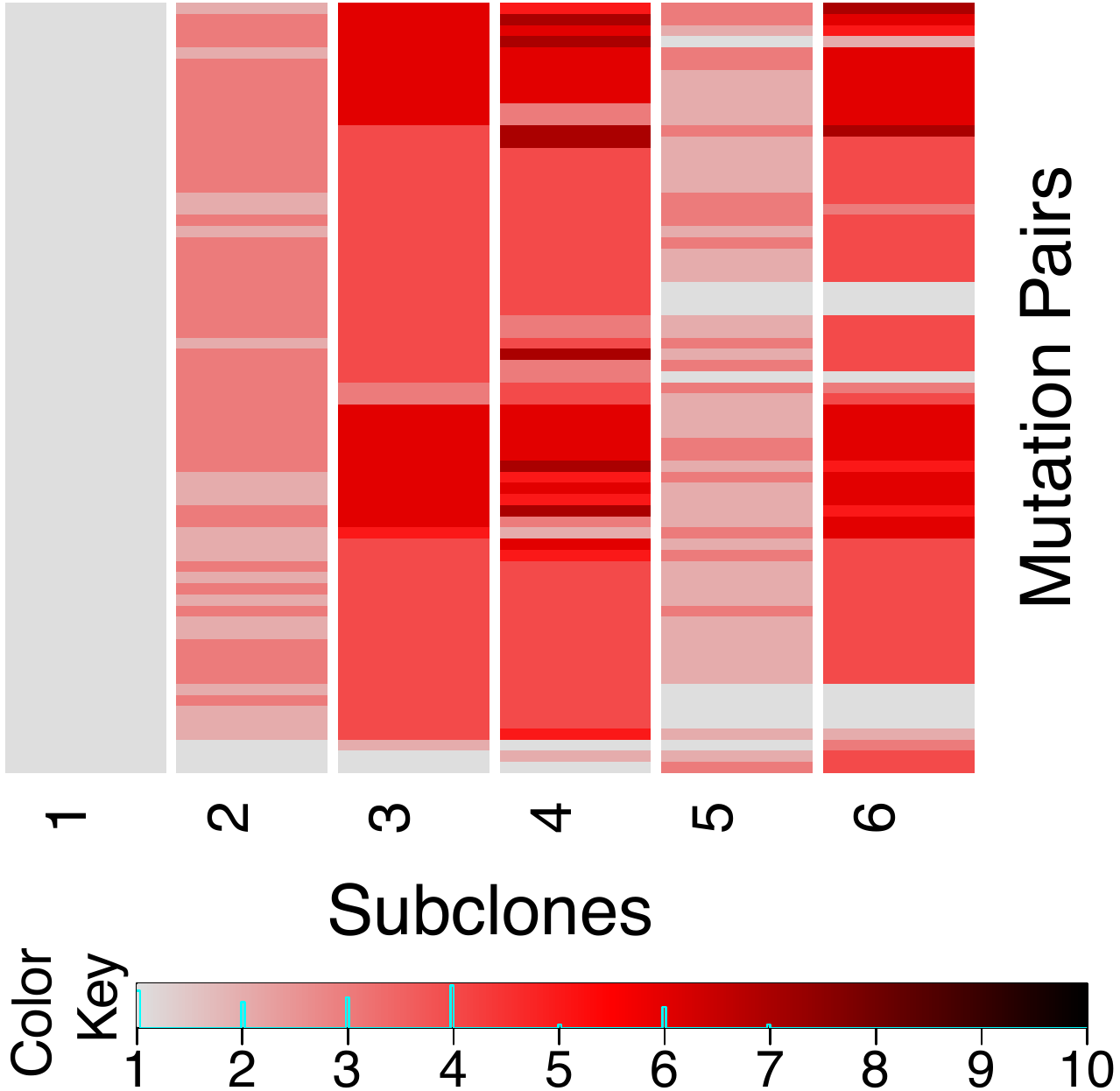}
\caption{$\Zhat$}
\end{subfigure}
\begin{subfigure}[t]{.325\textwidth}
\centering
\includegraphics[width=\textwidth]{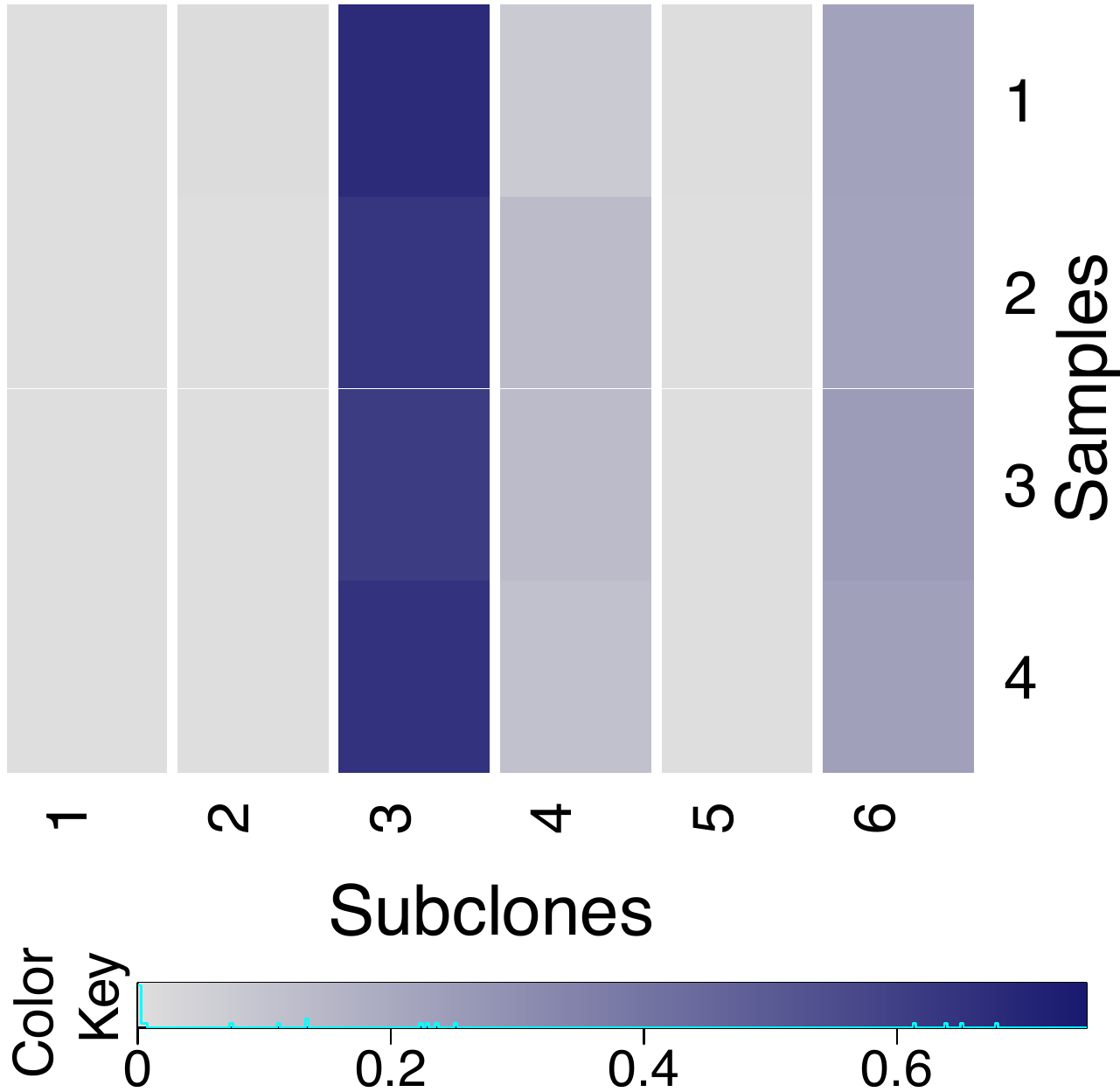}
\caption{$\what$}
\end{subfigure}
\begin{subfigure}[t]{.325\textwidth}
\centering
\includegraphics[width=\textwidth]{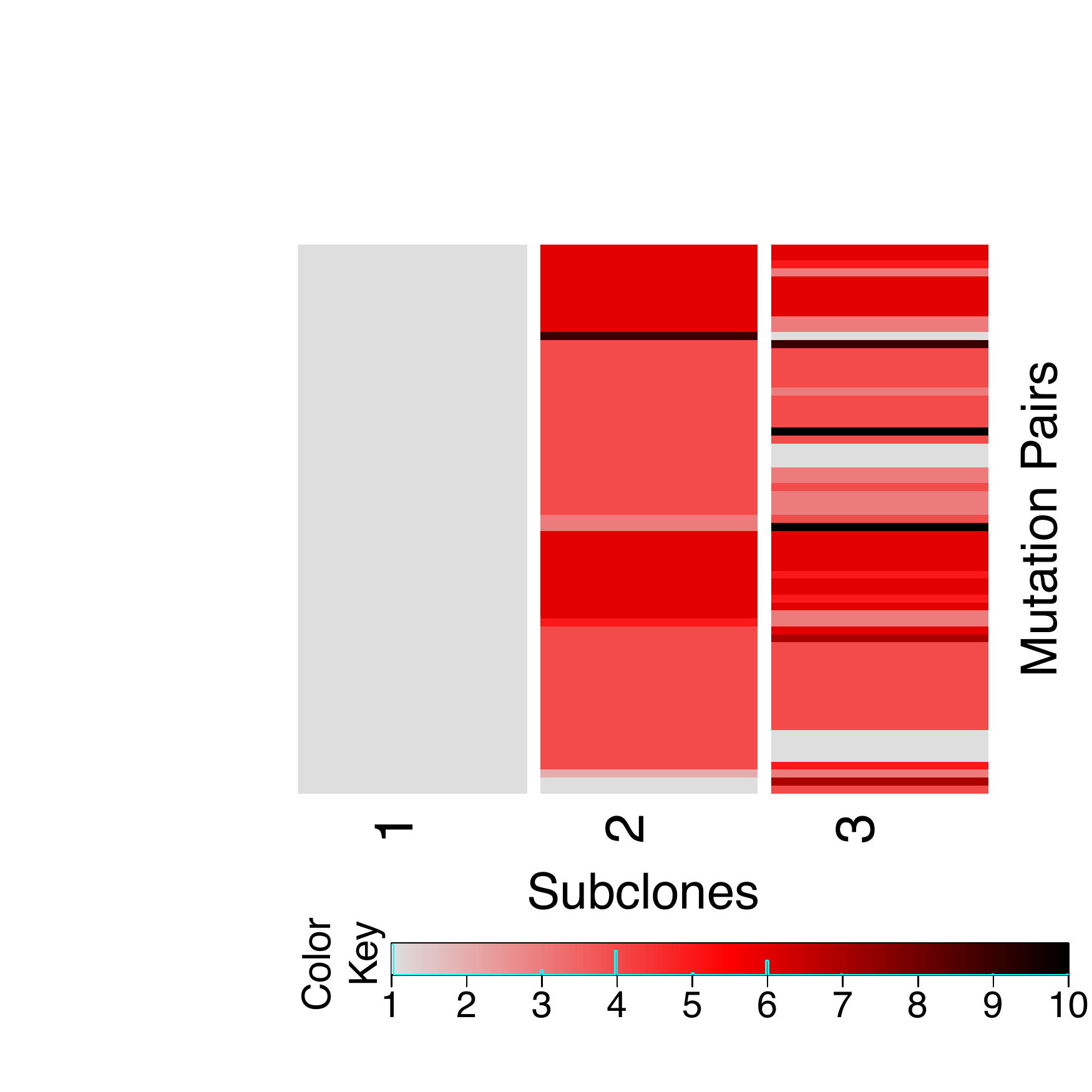}
\caption{$\Zhat^{\text{PairClone}}$}
\end{subfigure}
\end{center}
\caption{Simulation 3. Simulation truth $\bZ$ (a) and $\bw$ (b), and posterior inference under TreeClone (c, d, e) and  PairClone (f). }
\label{fig:sim2}
\end{figure}
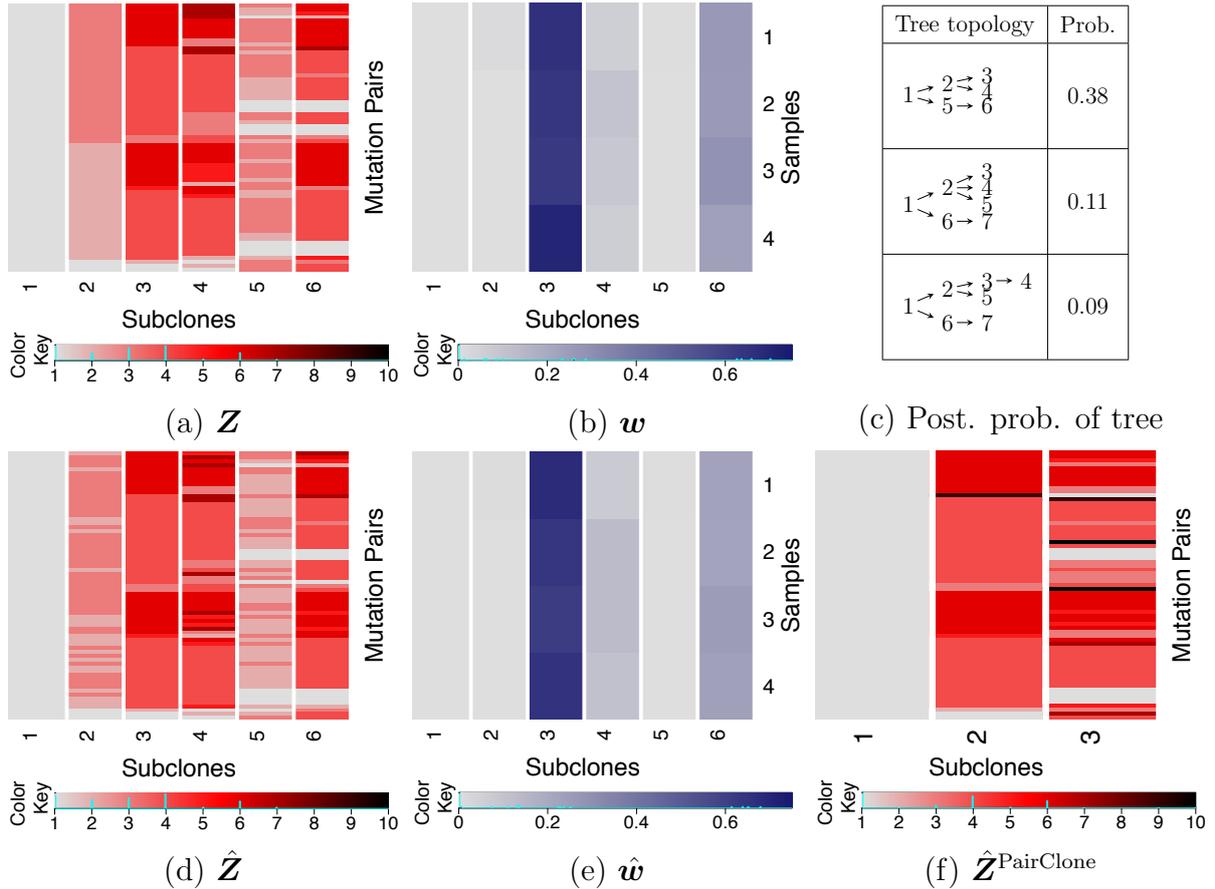

Fig. \ref{fig:sim2}(a, b) shows the simulation truth $\bZ$ and $\bw$ in
the form of heatmaps, respectively. We show $\bw$ with a light gray to
deep blue scale. A darker blue color indicates higher abundance of a
subclone in a sample
(the proportions $w_{t0}$ of the background subclone are tiny and not
shown). 
Read depths $\{ N_{tk} \}$ and left and right missing rates $\{
v_{tk2}, v_{tk3} \}$ are taken to be exactly the same as in the real
data. The average read depth is $\Nbar_{tk}=156$.

The hyperparameters are set as in Simulation 1. To
explore a larger tree space, we set $C_{\max} = 7$, run a total of
30000 MCMC iterations and discard the first 3000 draws as initial
burn-in.

Posterior inference is summarized in Fig.~\ref{fig:sim2}(c, d,
e). Fig.~\ref{fig:sim2}(c) shows the top three tree topologies and
corresponding posterior probabilities. The posterior mode recovers the
true phylogeny. Fig.~\ref{fig:sim2}(d, e) shows the estimated genotypes
$\Zhat$ and subclone proportions $\what$ conditional on $(\Chat,
\Tauhat)$. There are some mismatches for subclones 2, 4 and 5, as they
only take small proportions in all the samples ($w_{t2}, w_{t5} <
0.02$ and $w_{t4} < 0.11$ for all $t$). The reconstruction errors are
$Z_{\err} = 0.21$ and $w_{\err} = 0.01$ (considering only
subclones 1, 3 and 6 the reconstruction error becomes $Z_{\err}^{(1,
  3, 6)} = 0.04$). 

For comparison we again run Cloe, PhyloWGS and PairClone on the same data.
Cloe infers three subclones with phylogeny $ 1 \rightarrow 2
\rightarrow 3$, where Cloe's subclones 1, 2 and 3 roughly correspond
to the true subclones 1, 6 and 3, respectively. Cloe's result is
reasonable since subclones 2, 4 and 5 have small proportions and
 there is not much statistical power to estimate them. 
Similar to the definition of $Z_{\err}$, we define $Z_{\err, \text{ Cloe}}$ for
SNVs. Comparing $\Zhat^{\text{Cloe}}$ with subclones 1, 6 and 3 we
calculate the reconstruction error $Z_{\err, \text{ Cloe}}^{(1, 3, 6)}
= 0.04$, indicating good model fit of Cloe.  By allowing mutation
loss, Cloe infers a linear phylogenetic tree, which is still
reasonable.  On the other hand, PhyloWGS infers the phylogeny as
$
0 \rightarrow 1 \rightarrow 2 
\begin{array}{c}
\rightarrow 3  \\
\rightarrow 4 
\end{array}
$ (details not shown), which approximates but misses some detail in the simulation truth.
Finally, PairClone infers three subclones corresponding to the true subclones 1, 3 and 6, shown in Fig.~\ref{fig:sim2}(f). PairClone also reasonably recovers the truth but does not infer phylogeny. Comparing $\Zhat^{\text{PairClone}}$ with subclones 1, 3 and 6 we calculate the reconstruction error $Z_{\err, \text{ PairClone}}^{(1, 3, 6)} = 0.14$, which is higher than $Z_{\err}^{(1, 3, 6)}$.

\section{Lung Dataset}
\label{sec:real}
\begin{figure}[h!]
\begin{center}
\begin{subfigure}[t]{.325\textwidth}
\centering
\includegraphics[width=\textwidth]{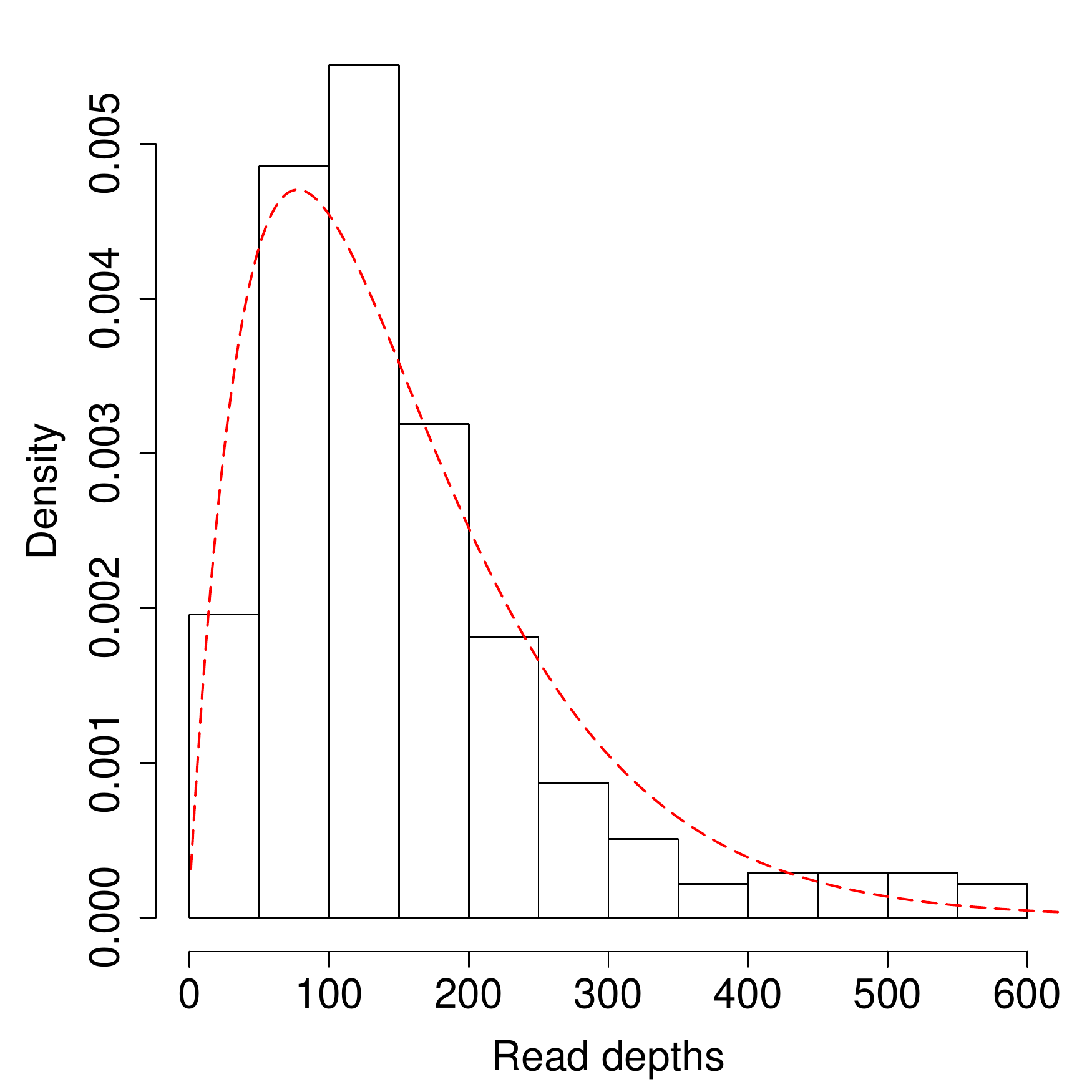}
\caption{Histogram of $\{ N_{tk}\}$}
\end{subfigure}
\begin{subfigure}[t]{.325\textwidth}
\centering
\includegraphics[width=\textwidth]{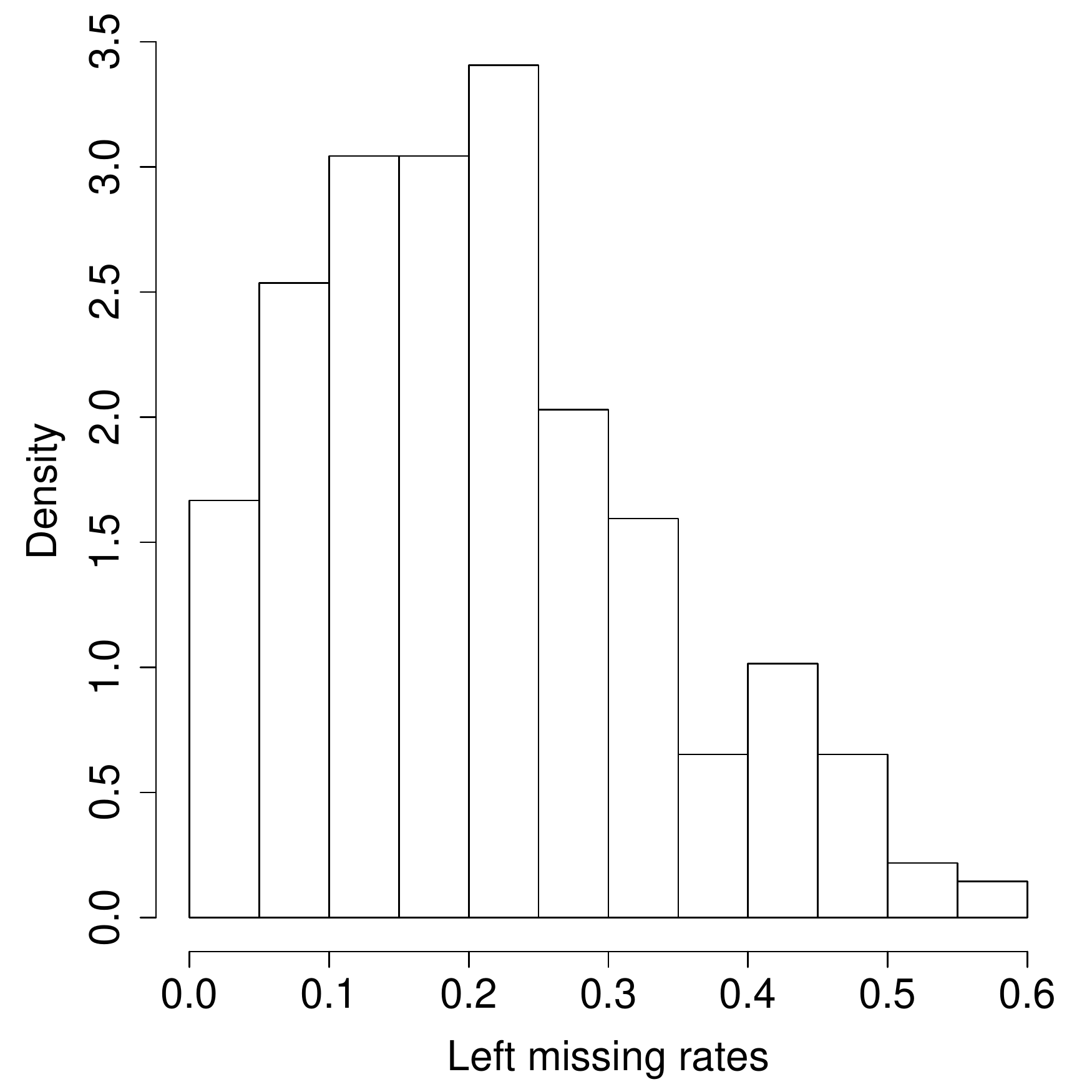}
\caption{Histogram of $\{ v_{tk2}\}$}
\end{subfigure}
\begin{subfigure}[t]{.325\textwidth}
\centering
\includegraphics[width=\textwidth]{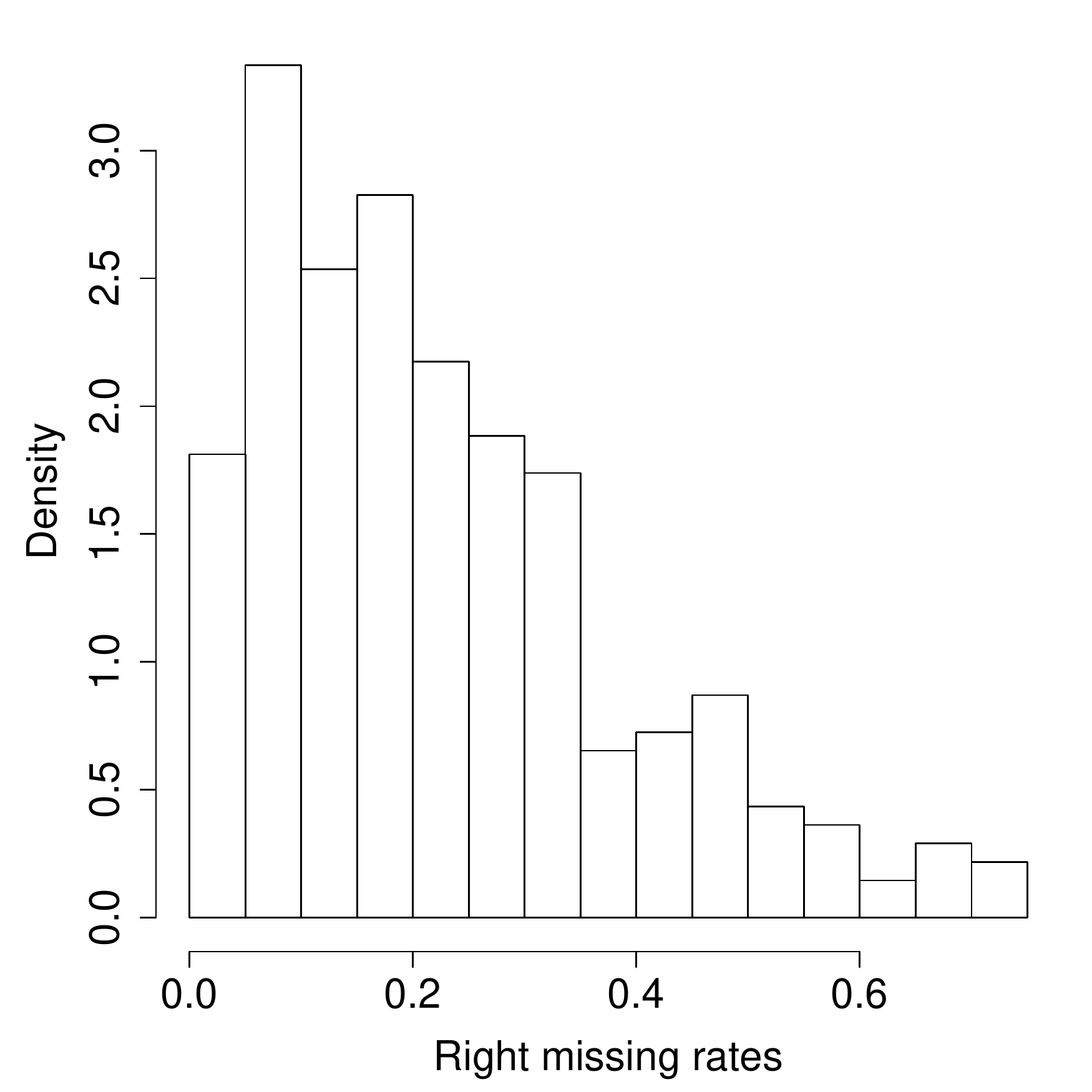}
\caption{Histogram of $\{ v_{tk3}\}$}
\end{subfigure}
\end{center}
\caption{Some summary plots of the lung cancer dataset. Histograms of read depths (a), left missing rates (b) and right missing rates (c). The red dashed line in (a) is a negative binomial density, showing that it is not unreasonable to assume the read depths follow a negative binomial distribution.  }
\label{fig:lung_data_summary}
\end{figure}

We use whole-exome sequencing (WES) data generated from four ($T = 4$)
surgically dissected tumor samples taken from a single patient
diagnosed with lung adenocarcinoma. DNA is extracted from all four
samples and the exome library is sequenced on an Illumina HiSeq 2000
platform in paired-end fashion. Each read is 100 base-pairs long. 
We use BWA [\cite{BWA}] and GATK's UniformGenotyper [\cite{GATK}] for
mapping and variant calling, respectively. In order to find mutation
pair locations along with their genotypes  and the  number of
reads supporting them, we use a bioinformatics tool called
\texttt{LocHap} [\cite{sengupta2016ultra}].
\texttt{LocHap} searches for two or
three SNVs that are scaffolded by the same reads. When the scaffolded
SNVs, known as local haplotypes, exhibit more than two haplotypes, it
is known as local haplotype variant (LHV). Using the individual BAM
and VCF files \texttt{LocHap} finds a few hundreds LHVs on  
average in a WES sample. We select LHVs with two SNVs as we are interested in mutation pairs only. On those LHVs, we run 
the bioinformatics filters suggested by \texttt{LocHap} to keep the mutation
pairs with high calling quality. 
We focus our analysis in copy number neutral regions. In the end, we
get 69 mutation pairs for the sample and record the read count data
from the output of \texttt{LocHap}. Figure \ref{fig:lung_data_summary}
shows the histograms of read depths, left missing rates and right
missing rates.  
The average read depth, left missing rate and right missing rate are
$156$, $0.21$ and $0.23$, respectively.
 Simulation 1 showed that with $T=4$ samples TreeClone should
provide useful inference with this combination of 
moderate read depth and left/right missing rates.

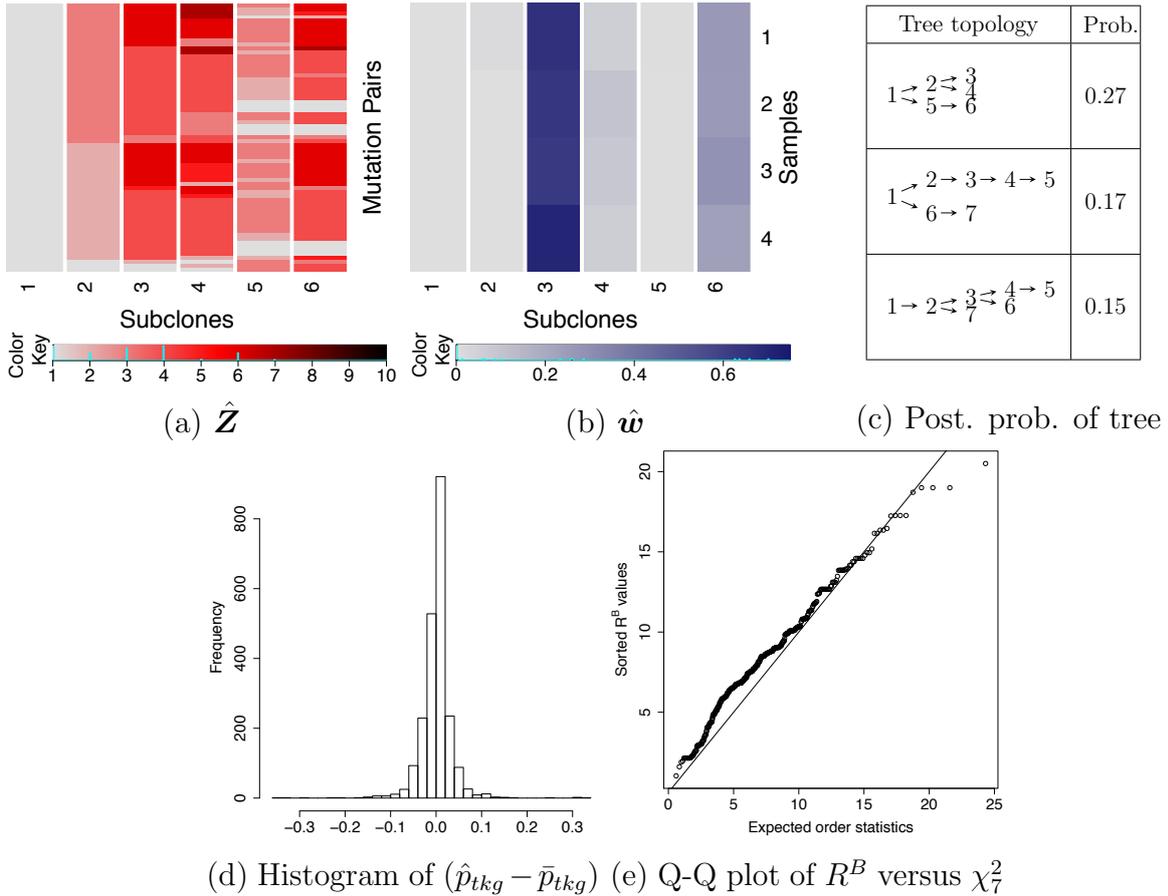
\begin{figure}[h!]
\begin{center}
\begin{subfigure}[t]{.325\textwidth}
\centering
\includegraphics[width=\textwidth]{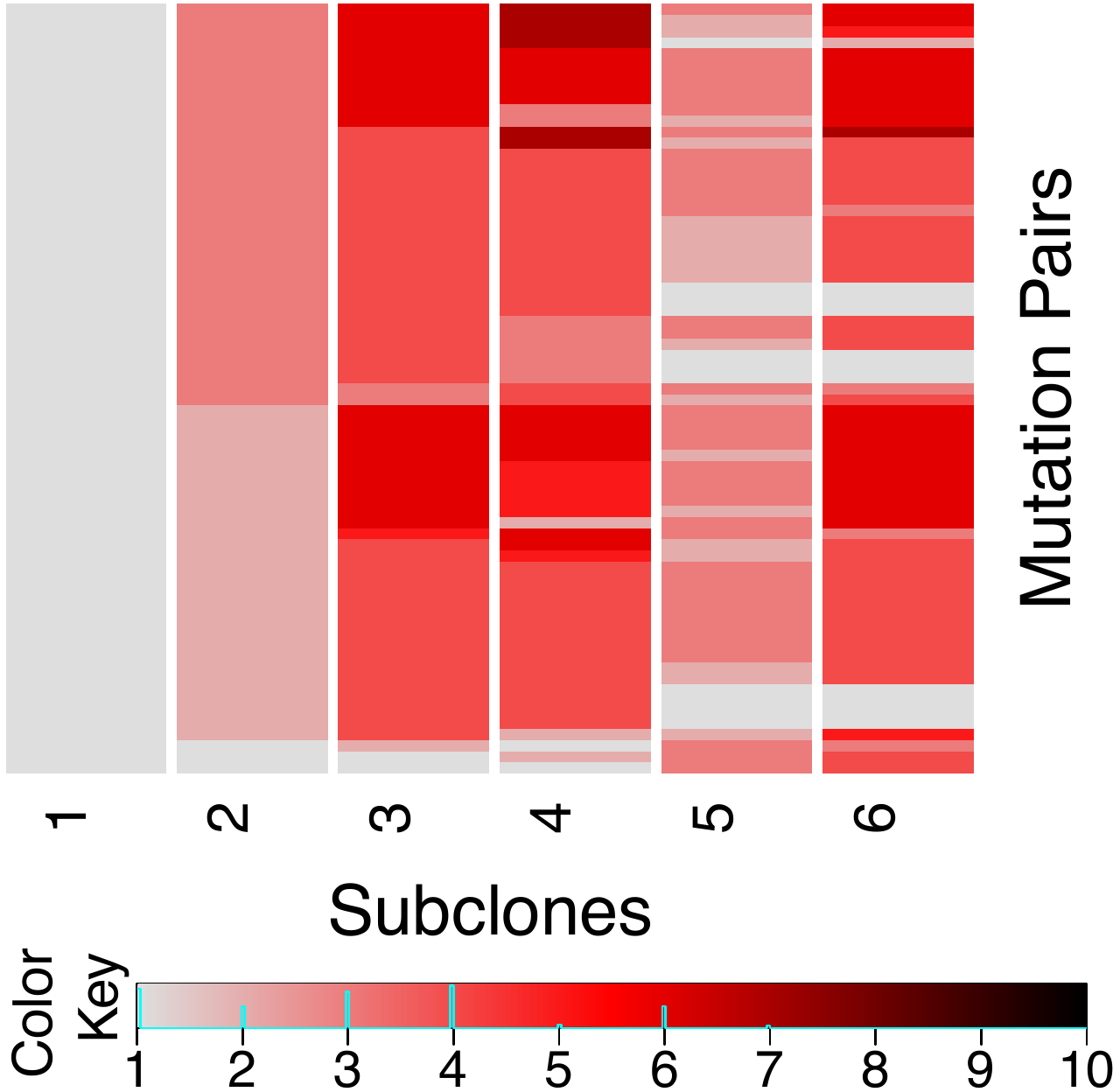}
\caption{$\Zhat$}
\end{subfigure}
\begin{subfigure}[t]{.325\textwidth}
\centering
\includegraphics[width=\textwidth]{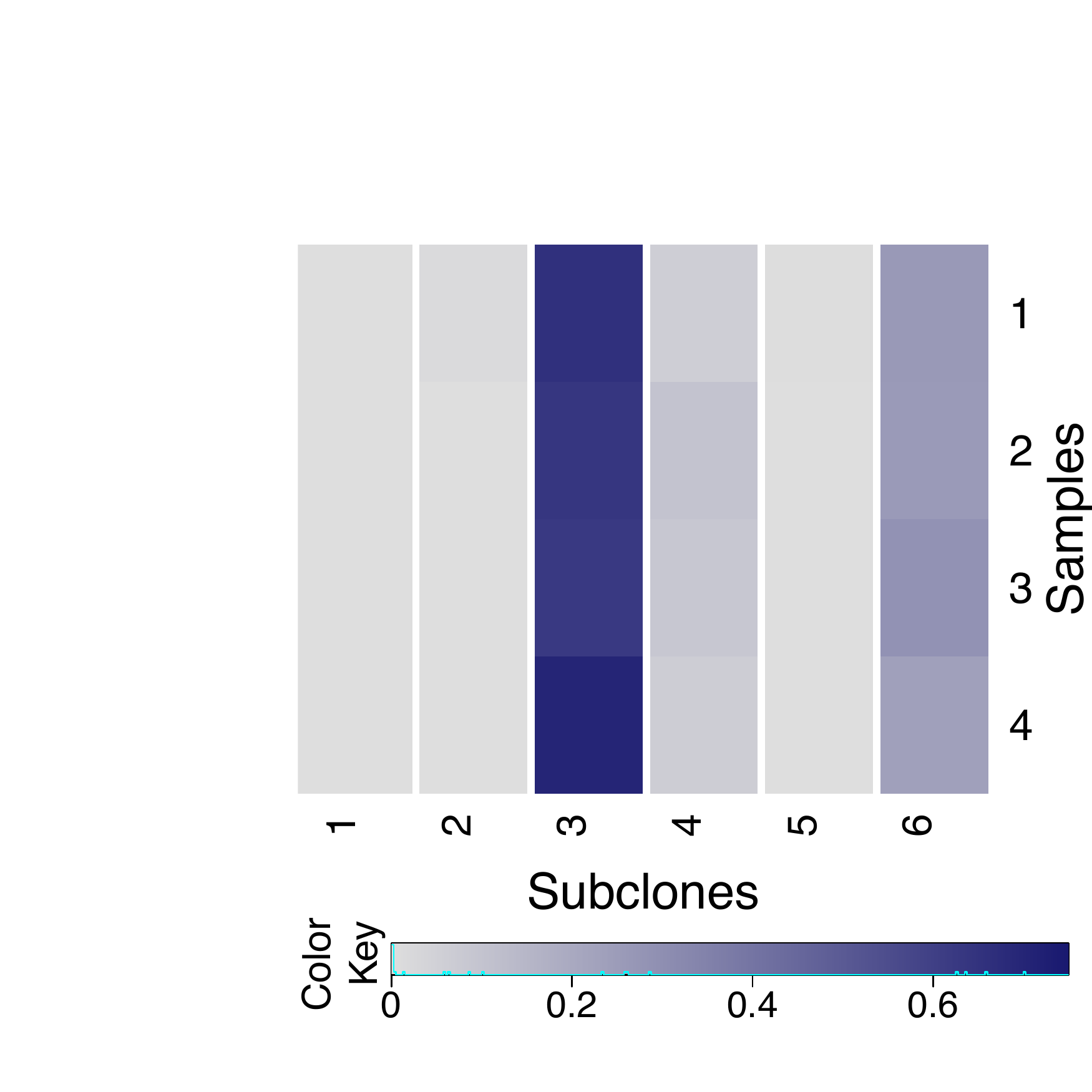}
\caption{$\what$}
\end{subfigure}
\begin{subfigure}[t]{.325\textwidth}
\centering
\vspace{-50mm}
\scalebox{0.74}{
\begin{tabular}{|>{\raggedright\arraybackslash}m{32mm}|>{\centering\arraybackslash}m{8mm}|@{}m{0pt}@{}}
\hline
\multicolumn{1}{|c|}{Tree topology} & Prob. & \\ \hline
\begin{tikzpicture}[grow=right, sloped]
\tikzstyle{level 1}=[level distance=0.7cm, sibling distance=0.4cm]
\tikzstyle{level 2}=[level distance=0.7cm, sibling distance=0.3cm]
\node[bag] {1}
    child {
        node[bag] {5}
            child {
                node[bag] {6}
                edge from parent [-stealth]
            }
            edge from parent [-stealth]
    }
    child {
        node[bag] {2}        
        child {
                node[bag] {4}
                edge from parent [-stealth]
            }
            child {
                node[bag] {3}
                edge from parent [-stealth]
            }
        edge from parent [-stealth]       
    };
\end{tikzpicture} & 0.27 &  \\[3em]
\hline
\begin{tikzpicture}[grow=right, sloped]
\tikzstyle{level 1}=[level distance=0.7cm, sibling distance=0.6cm]
\tikzstyle{level 2}=[level distance=0.7cm, sibling distance=0.3cm]
\node[bag] {1}
    child {
        node[bag] {6}
            child {
                node[bag] {7}
                edge from parent [-stealth]
            }
            edge from parent [-stealth]
    }
    child {
        node[bag] {2} 
        child {
                node[bag] {3}
                child {
                  node[bag] {4}
                  child {
                    node[bag] {5}
                    edge from parent [-stealth]
                  }
                  edge from parent [-stealth]
                }
                edge from parent [-stealth]
            }       
        edge from parent [-stealth]       
    };
\end{tikzpicture}
 & 0.17  & \\[3em]
 \hline
\begin{tikzpicture}[grow=right, sloped]
\tikzstyle{level 1}=[level distance=0.7cm, sibling distance=0.6cm]
\tikzstyle{level 2}=[level distance=0.7cm, sibling distance=0.3cm]
\node[bag] {1}
    child {
        node[bag] {2}
            child {
                node[bag] {7}
                edge from parent [-stealth]
            }
            child {
                node[bag] {3}
                child {
                  node[bag] {6}
                  edge from parent [-stealth]
                }
                child {
                  node[bag] {4}
                  child {
                    node[bag] {5}
                    edge from parent [-stealth]
                  }
                  edge from parent [-stealth]
                }
                edge from parent [-stealth]
            }
            edge from parent [-stealth]
    };
\end{tikzpicture}
 & 0.15 &  \\[3em]
 \hline
\end{tabular}
}
\vspace{2.5mm}
\caption{Post. prob. of tree}
\end{subfigure}
\begin{subfigure}[t]{.325\textwidth}
\centering
\includegraphics[width=\textwidth]{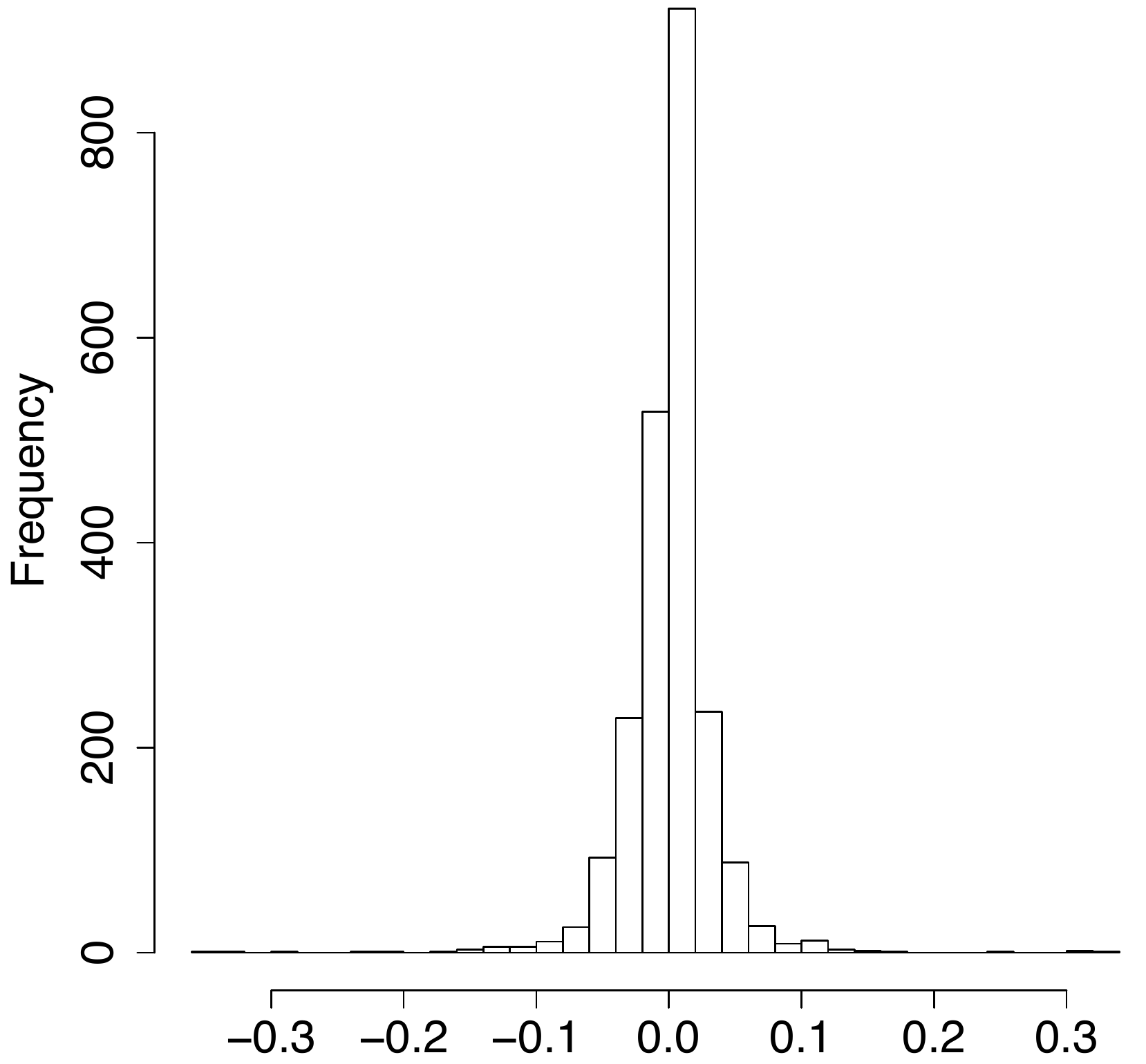}
\caption{Histogram of $(\phat_{tkg} - \bar{p}_{tkg})$}
\end{subfigure}
\begin{subfigure}[t]{.325\textwidth}
\centering
\includegraphics[width=\textwidth]{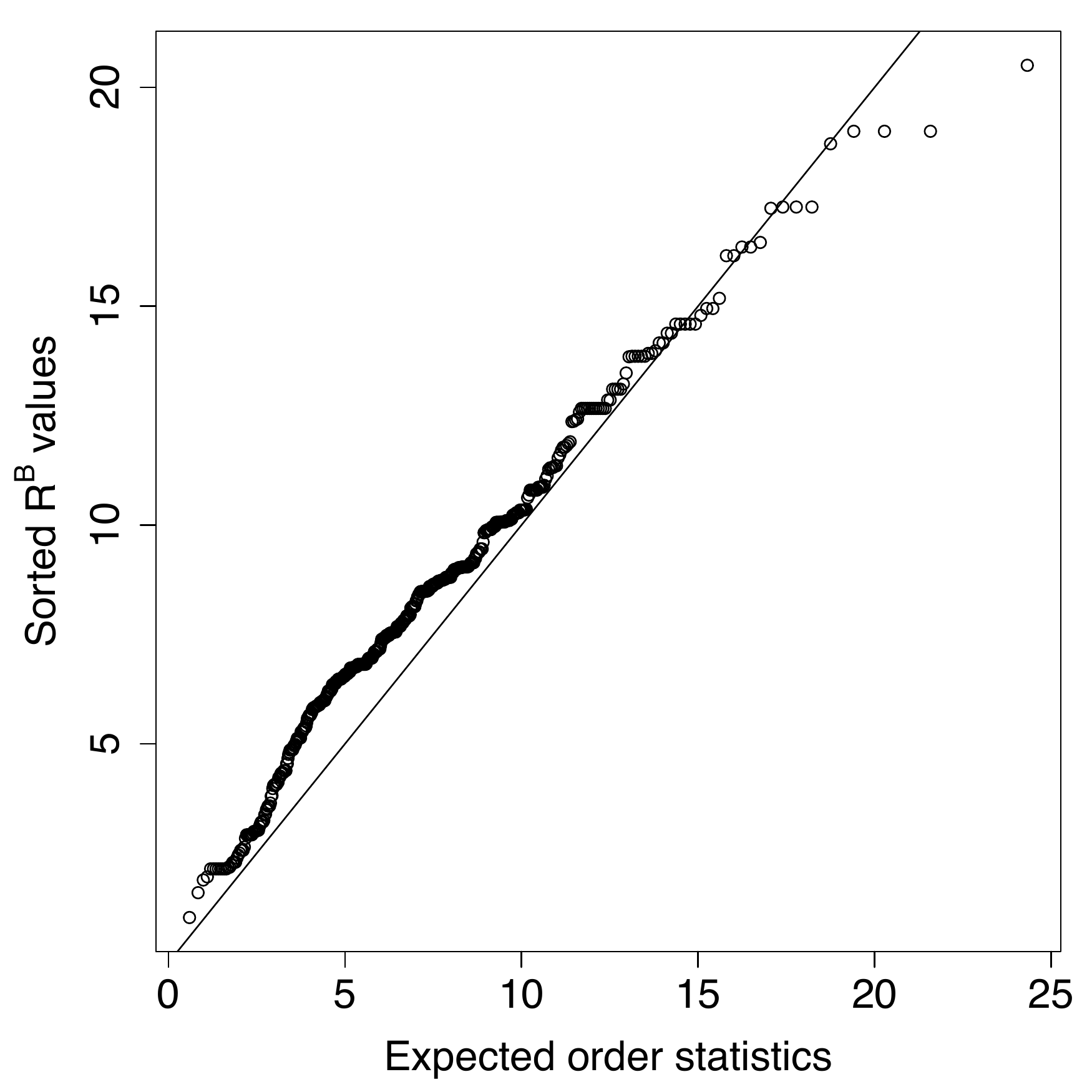}
\caption{Q-Q plot of $R^B$ versus ${\chi}_7^2$}
\end{subfigure}
\end{center}
\caption{Posterior inference with TreeClone for lung cancer data set.}
\label{fig:lung}
\end{figure}

We use the same hyperparameters as in Simulation 1. To explore a larger tree space, we set $C_{\max} = 7$, run a total of 30000 MCMC iterations and discard the first 3000 draws as initial burn-in. 
Fig. \ref{fig:lung}(c) shows 
the top three tree topologies and corresponding posterior probabilities. The posterior mode is
\begin{center}
\begin{tikzpicture}[grow=right, sloped]
\node[bag] {1}
    child {
        node[bag] {5}
            child {
                node[bag] {6}
                edge from parent [-stealth]
            }
            edge from parent [-stealth]
    }
    child {
        node[bag] {2}        
        child {
                node[bag] {4}
                edge from parent [-stealth]
            }
            child {
                node[bag] {3}
                edge from parent [-stealth]
            }
        edge from parent [-stealth]       
    };
\end{tikzpicture}
\end{center}
with $C=6$
subclones. Fig. \ref{fig:lung}(a, b) show the estimated subclone
genotypes $\Zhat$ and cellular proportions $\what$, respectively ($\hat{w}_{t0} < 3 \times 10^{-3}$ are not shown). The
rows for $\Zhat$ are reordered for better display. The cellular
proportions of the subclones show strong similarity across the 4
samples, indicating homogeneity of the samples. 
This is plausible as the samples are dissected from proximal
sites. Subclone 1, 
which is the normal subclone, takes a small proportion in all 4
samples, indicating high purity of the tumor samples. Subclones 2
and 5 are also included in only small proportions.
They have almost vanished in the samples. 
However, as parents of subclones 3, 4 and 6, respectively, they are
important for the reconstruction of the subclone phylogeny.
Subclones 3, 4 and 6 are the three main subclones. They
share a large proportion of common mutations, but each one has some
private mutations, consistent with the estimated tree.

\paragraph{Test of fit}
Finally,  Fig. \ref{fig:lung}(d) shows a histogram
of residuals, where we calculate empirical values $\bar{p}_{tkg} =
n_{tkg} / N_{tk}$ and plot the difference $(\phat_{tkg} -
\bar{p}_{tkg})$. The residuals are centered at zero with little
variation, indicating a good model fit. 
For a more formal goodness-of-fit test, we carried out the Bayesian
${\chi}^2$ test proposed in \cite{johnson2004bayesian}. Recall that the
observations 
in our case are short reads $\bs_{tki}$ taking $G = 8$ discrete values
$\{00, 01, 10, 11, -0, -1, 0-, 1- \}$. We count the number of short
reads that fall into each of these categories. Let $M_g$ denote these
counts, and let $\bx^{(l)}$ be a posterior sample of $\bx = (\bZ, \bw,
\brho)$. The statistic $R^B$ is defined as 
\begin{align*}
R^B(\bx^{(l)}) = \sum_{g=1}^G \left[ \frac{M_g - Nq_g(\bx^{(l)})}{\sqrt{Nq_g(\bx^{(l)})}}\right]^2,
\end{align*}
where $N = \sum_{t, k} N_{tk}$ and $q_g(\bx^{(l)}) = \sum_{t, k}
N_{tk} p_{tkg}(\bx^{(l)}) / N$ is the expected proportion of short
reads in category $g$ calculated by $\bx^{(l)}$.
 Under the null hypothesis of a good model fit the statistic
should follow a $\chi^2$-distribution with $G-1 = 7$ degrees of
freedom. 
Fig. \ref{fig:lung}(e) shows a
quantile-quantile plot of posterior samples of $R^B$ against expected
order statistics from a ${\chi}_7^2$ distribution. In addition, find
the proportion of posterior samples of $R^B$ exceeding the
$95\%$ quantile of a  ${\chi}_7^2$ distribution to be 0.054. There
is no evidence of a lack of fit. 

\begin{figure}[h!]
\begin{center}
\begin{subfigure}[t]{.25\textwidth}
\centering
\includegraphics[width=\textwidth]{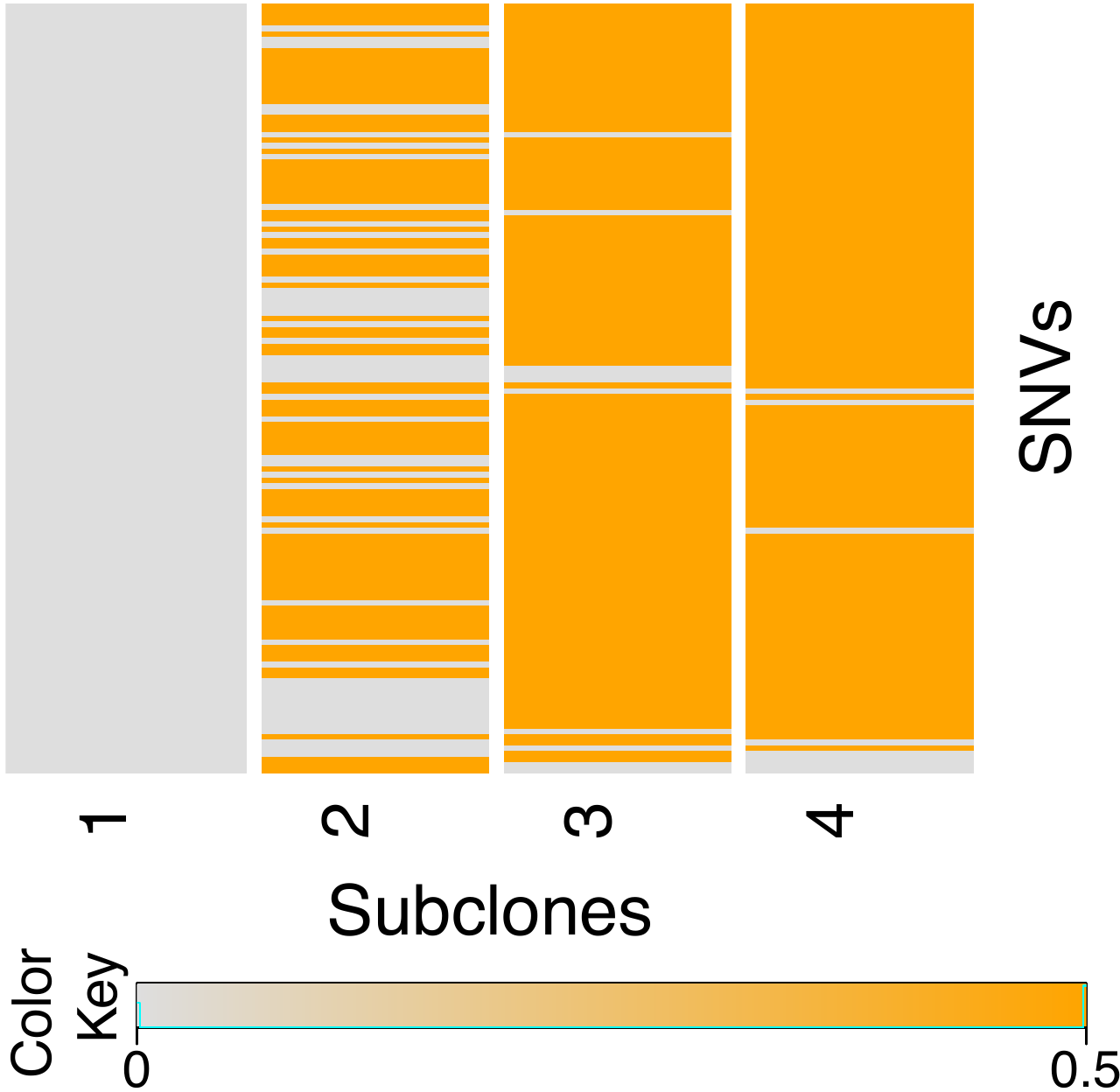}
\caption{$\Zhat^{\text{Cloe}}$}
\end{subfigure}
\begin{subfigure}[t]{.25\textwidth}
\centering
\includegraphics[width=\textwidth]{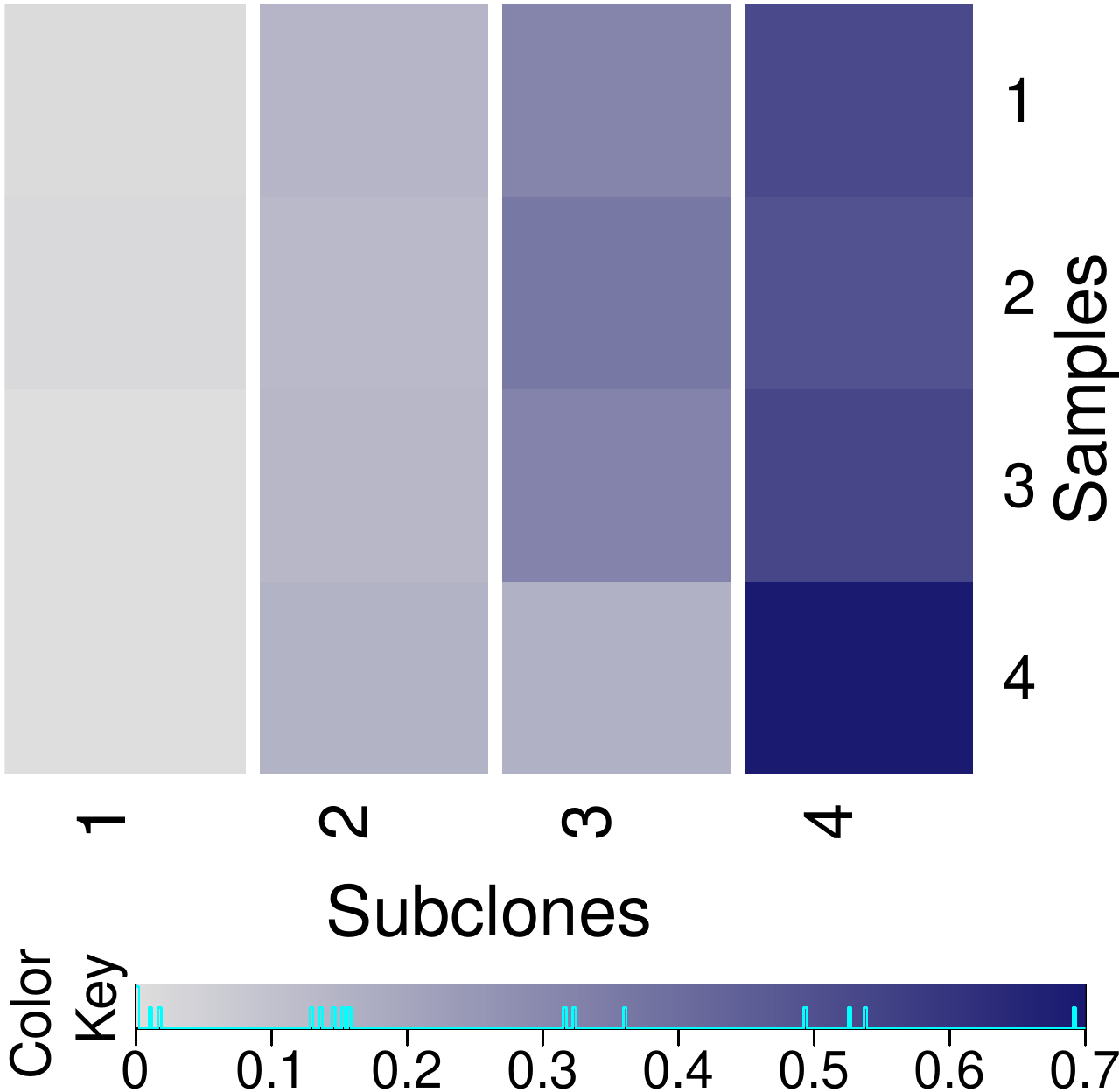}
\caption{$\what^{\text{Cloe}}$}
\end{subfigure}
\hspace{1mm}
\begin{subfigure}[t]{.45\textwidth}
\centering
\vspace{-40mm}
\scalebox{0.62}{
\begin{tabular}{|l|c|c|c|c|}
\hline
\backslashbox{Clusters \\ (sizes)}{Samples} & 1 & 2 & 3 & 4 \\\hline
\qquad \quad 0 (0) & 1.0 & 1.0 & 1.0 & 1.0  \\\hline
\qquad \quad 1 (93) & 0.977 & 0.990 &  0.989 & 0.980 \\\hline
\qquad \quad 2 (32) & 0.854 & 0.816 & 0.832 & 0.852 \\\hline
\qquad \quad 3 (6) & 0.532 & 0.431 & 0.473 & 0.596 \\\hline
\qquad \quad 4 (3) & 0.433 & 0.408 & 0.445 & 0.386 \\\hline
\qquad \quad 5 (3) & 0.252 & 0.180 &  0.229 & 0.152 \\\hline
\qquad \quad 6 (1) & 0.112 & 0.108 & 0.056 & 0.127 \\\hline
\end{tabular}
}
\vspace{1.4mm}
\caption{Cellular prevalence (PhyloWGS)}
\end{subfigure}
\end{center}
\caption{Posterior inference with Cloe (a, b) and PhyloWGS (c) for lung cancer data set.}
\label{fig:lung_compare}
\end{figure}

\paragraph{ Cloe and PhyloWGS}
For comparison, we run Cloe 
and PhyloWGS on the same data set with default hyperparameters.
Cloe infers four subclones with phylogeny $ 1 \rightarrow 2 \rightarrow 3 \rightarrow 4$. 
Fig. \ref{fig:lung_compare}(a, b) show the estimated genotypes $\Zhat^{\text{Cloe}}$ and cellular proportions $\what^{\text{Cloe}}$, respectively. 
The estimated subclones 2, 3 and 4 under Cloe match with subclones 6,
4 and 3, respectively, under TreeClone. Cloe infers
a linear phylogenetic tree since it allows mutation loss.  
PhyloWGS estimates 6 clusters (and a cluster 0 for normal subclone) of
the SNVs with phylogeny 
\begin{center}
\begin{tikzpicture}[grow=right, sloped]
\node[bag] {0}
    child {
        node[bag] {1}
            child {
                node[bag] {2}
                child {
                  node[bag] {5}
                  child {
                    node[bag] {6.}
                    edge from parent [-stealth]
                  }
                  edge from parent [-stealth]
                }
                child {
                  node[bag] {3}
                  child {
                    node[bag] {4}
                    edge from parent [-stealth]
                  }
                  edge from parent [-stealth]
                }
                edge from parent [-stealth]
            }
            edge from parent [-stealth]
    };
\end{tikzpicture}
\end{center}
Fig. \ref{fig:lung_compare}(c) summarizes the cluster sizes and
cellular prevalences. In light of the earlier simulation studies we
believe that the inference on $\Tau$ under TreeClone is more reliable. 
However, results from Cloe and PhyloWGS confirm that the four samples
have similar proportions  
of all the subclones, indicating little inter-sample heterogeneity.  
Also, Cloe and PhyloWGS infer very small normal cell proportion,
 confirming the high tumor purity found by TreeClone. 
Finally, the same lung cancer dataset was analyzed under
PairClone in \cite{zhou2017pairclone}. PairClone infers three
subclones corresponding to TreeClone's subclones 1, 3 and 6 and also
confirms the 4 tissue samples are highly homogenuous. PairClone gives
reasonable result but can not infer phylogeny.

\section{Discussion}
\label{sec:disc}

In this work, using a tree-based LFAM we infer subclonal
genotypes structure for 
mutation pairs, their cellular proportions and the phylogenetic
relationship among subclones. This is the first attempt to generate a
subclonal phylogenetic structure using mutation pair data.
We show 
that more accurate inference can be obtained using mutation
pairs data compared to using only marginal counts for single SNVs. 
The model can be easily extended to incorporate more than two
SNVs. Another way of extending the model is to encode mutation times
inside the length of the edges of phylogenetic tree.
 
The major motivation for accurate estimation of heterogeneity in tumor
is personalized medicine. The next step towards this goal is to
use varying estimates of subclonal proportions across patients to
drive adaptive treatment allocation.

Currently the heterogeneity is measured
mostly with SNV and CNA data. However, structural variants (SVs) like
deletion, duplication, inversion, translocation and other large genome
rearrangement arguably provide more 
accurate  data for VAF estimation  [\cite{fan2014towards}],
which is the key input for characterizing  tumor 
heterogeneity. Therefore incorporation of SVs into the current model
could significantly improve the outcome of tumor heterogeneity
analysis.
Recently, in~\cite{brocks2014intratumor} the authors 
attempted to explain intratumor
heterogeneity in DNA methylation and copy-number pattern by a unified
evolutionary process. The current genome centric definition of
tumor heterogeneity could be extended by incorporation of methylation,
DNA mutation, and RNA expression data in an integromics model.

Finally in the era of big data it is important to factor computation
into the research effort, and build 
efficient computational models that could handle millions of
SNVs. Linear response variational Bayes [\cite{giordano2015linear}] or
MAD-Bayes [\cite{broderick2013mad,xu2015mad}] methods could be
considered as alternative computational strategies to tackle the
problem.

\section*{Appendix}
\subsection*{Glossary of terms}

\paragraph{SNV}
A single nucleotide variant or SNV is a DNA sequence variation where one nucleotide is replaced by anther nucleotide. This is  very similar to a single nucleotide polymorphism or SNP. In our paper, the term SNV is preferred over SNP because the latter includes an additional interpretation about variants in a population. This means that an SNV could potentially be a SNP but this cannot be determined at the point where the variant is detected in a single sample.

\paragraph{SCNA}
Copy number aberration (CNA) is gain or loss of large segments of the genome ranging in size from a few kilobases to a whole chromosomes. Somatic CNA (SCNA) that occurs during the lifetime of an individual is a major contributor to cancer development, particularly for solid tumors.

\paragraph{Short-read}
Short DNA sequences produced by the sequencing machine. The range of the read length of a short-read sequencing instrument is between $100$ and $600$ bp.

\paragraph{Phasing}
Phasing helps to identify the copy of the chromosome (paternal or  maternal chromosome) to which a particular allele belongs or alternatively, which alleles appear together on the same chromosome. In short-read sequencing, for example, it is difficult to resolve the haplotype of two heterozygous SNPs if they have not been covered by the same read. If you observe A/C and G/T, it is difficult to know whether we have AG and CT or CG and AT.

\paragraph{VAF}
Variant allele frequency or VAF is the relative frequency of an alternative allele at a particular genomic locus expressed as a fraction or percentage. The quantity is calculated as the fraction of short-reads overlapping a genomic locus that support the non-reference (mutant/alternate) allele.

\paragraph{Purity}
Tumor purity is an essential component in finding tumor heterogeneity. The tumor micro-environment contains other non-cancerous cell types including fibroblasts, immune cells, endothelial cells and normal epithelial cells along with cancer cells. Tumor purity is the proportion of cancer cells in the sample. 

\paragraph{Phylogenetic tree}
A phylogenetic tree of a tumor sample is a branching diagram or ``tree" showing the evolutionary relationships among various subclones that co-exist in the sample. 

\paragraph{Subclone calling method}
Subclone calling method refers to a computation algorithm that is used to infer the subclonal structure in a tumor sample. 


\bibliographystyle{Chicago}
\bibliography{ref_treeclone}

\end{document}